\documentclass[namedreferences]{solarphysics}
\usepackage[optionalrh]{spr-sola-addons} 
\usepackage{graphicx}        
\usepackage{color}           
\usepackage{url}             
\usepackage{ulem}
\usepackage{pdflscape}
\newcommand{\etal}{{\it et al.}}
\newcommand{\eegg}{{\it e.g. }}
\newcommand{\iiee}{{\it i.e. }}
\newcommand{\myAlfvenic}{{Alfv\'enic }}
\newcommand{\myAlfven}{{Alfv\'en }}
\newcommand{\myinsitu}{{\it in-situ }}
\newcommand{\mysun}{\odot}%
\newcommand{\mykmps}{{${\rm km\,s}^{-1}$}}%
\newcommand{\mypcc}{{${\rm cc}^{-1}$}}%
\newcommand{\mykelvin}{{${\rm K}^{\circ}$}}%
\newcommand{\mysubzero}{{\rm 0}}%
\newcommand{\mysubB}{{\rm B}}%
\newcommand{\mysubb}{{\rm b}}%
\newcommand{\mysubc}{{\rm c}}%
\newcommand{\mysubi}{{\rm i}}%
\newcommand{\mysubj}{{\rm j}}%
\newcommand{\mysubk}{{\rm k}}%
\renewcommand{\vec}[1]{{\mathbfit #1}}
\newcommand{\aap}{    {\it Astron. Astrophys.}}

\newcommand{\apj}{    {\it Astrophys. J.}}
\newcommand{\apjl}{   {\it Astrophys. J. Lett.}}
\newcommand{\apjss}{  {\it Astrophys. J. Suppl. Ser.}}
\newcommand{\apss}{   {\it Astrophys. Space Sci.}}

\newcommand{\grl}{    {\it Geophys. Res. Lett.}}

\newcommand{\jcp}{    {\it J. Comp. Phys.}}
\newcommand{\jgr}{    {\it J. Geophys. Res.}}

\newcommand{\solphys}{{\it Solar Phys.}}

\begin{document}
\begin{article}
\begin{opening}
\title{The {\it Helioseismic and Magnetic Imager} (HMI)
Vector Magnetic Field Pipeline:
Magnetohydrodynamics Simulation Module for the Global Solar Corona}
\author{%
K.~\surname{Hayashi}$^{1,2}$
\sep
J.T.~\surname{Hoeksema}$^{1}$
\sep
Y.~\surname{Liu}$^{1}$
\sep
M.G.~\surname{Bobra}$^{1}$
\sep
X.D.~\surname{Sun}$^{1}$
\sep
A.A.~\surname{Norton}$^{1}$
}%
\runningauthor{K.Hayashi \etal}
\runningtitle{MHD Simulation in HMI Data Pipeline}
\institute{$^{1}$%
W. W. Hansen Experimental Physics Laboratory, Stanford University, CA, USA,
email:
\url{keiji@sun.stanford.edu}\\
\url{todd@sun.stanford.edu}\\ 
\url{yliu@sun.stanford.edu}\\
\url{mbobra@sun.stanford.edu}\\
\url{xudong@sun.stanford.edu}\\
\url{norton@sun.stanford.edu}
}%
\institute{$^{2}$ National Astronomical Observatory of China, Chinese Academy of Sciences, Beijing, China}
\begin{abstract}
Time-dependent three-dimensional magnetohydrodynamics (MHD) simulation modules are implemented at the Joint Science Operation Center (JSOC) of Solar Dynamics Observatory (SDO).
The modules regularly produce three-dimensional data of the time-relaxed minimum-energy state of the solar corona using global solar-surface magnetic-field maps created from Helioseismic Magnetic Imager (HMI) full-disk magnetogram data.
With the assumption of polytropic gas with specific heat ratio of 1.05, three types of simulation products are currently generated: i) simulation data with medium spatial resolution using the definitive calibrated synoptic map of the magnetic field with a cadence of one Carrington rotation, ii) data with low spatial resolution using the definitive version of the synchronic frame format of the magnetic field, with a cadence of one day, and iii) low-resolution data using near-real-time (NRT) synchronic format of the magnetic field on daily basis.
The MHD data available in the JSOC database are three-dimensional, covering heliocentric distances from 1.025 to 4.975 solar radii, and contain all eight MHD variables: the plasma density, temperature and three components of motion velocity, and three components of the magnetic field.
This article describes details of the MHD simulations as well as the production of the input magnetic-field maps, and details of the products available at the JSOC database interface.
In order to assess the merits and limits of the model, we show the simulated data in early 2011 and compare with the actual coronal features observed by the Atmospheric Imaging Assembly (AIA) and the near-Earth \myinsitu data.
\end{abstract}
%
\end{opening}
%
%
\section{Introduction}\label{intro} 
The Joint Science Operation Center (JSOC) data pipeline currently processes various types of observations made by the Helioseismic Magnetic Imager (HMI) instrument (\eegg \opencite{Scherrer12} and \opencite{Schou12}) and Atmospheric Imaging Assembly (AIA) instrument (\eegg \opencite{Lemen12}) onboard Solar Dynamics Observatory (SDO).
From the HMI observations, the data pipeline rapidly produces various types of magnetic field data (see \opencite{Liu12}; \opencite{Hoeksema14}; \opencite{Bobra14} and \opencite{Centeno14}) as well as helioseismology data and derivative products (\opencite{Zhao12}).
Details of the data production processes for the line-of-sight magnetograms and vector magnetic field are described by \inlinecite{Liu12} and \inlinecite{Hoeksema14}, respectively.

HMI observes the full-disk Sun continuously and almost seamlessly in time, with one-arcsecond spatial resolution and rapid temporal cadence, which allows us to generate frequent whole-Sun maps, like the synoptic map, representing the solar-surface condition on a time scale of one solar rotation with fine spatial scale.

The HMI observables do not include the solar corona (above the photosphere) which is not directly measured.
For various purposes, specifically for space weather, the theoretically determined three-dimensional solar corona and solar wind that depend on the magnetic field measured in the solar photosphere are crucially important.

As demonstrated by pioneering works, (\iiee \opencite{Wu83}; \opencite{Linker90}; \opencite{Dryer91}; \opencite{Washimi93}; \opencite{Usmanov93}), a time-dependent magnetohydrodynamics (MHD) simulation can provide us with powerful capability to inspect the complex nonlinear state of the global trans-\myAlfvenic solar corona and solar wind.
Specifically, by using the global solar-surface magnetic field map as an input to specify the boundary values, the MHD solution can represent the state of the solar corona at the time of observation (\eegg \opencite{Usmanov93}; \opencite{Usmanov95}).
Using time-relaxation MHD simulations, the simulated MHD variables interact with each other until the system reaches the minimum-energy state of the sub- and trans-\myAlfvenic solar corona and super-Alfvenic solar wind.
This solution can be regarded as the most plausible quiescent state of the global solar atmosphere.
The large-scale steady state is a cradle within which the magnetic energy in the streamer or small-scale active regions evolves and accumulates, and it defines a background state where interplanetary disturbance propagates.
As such, it is beneficial to the solar physics community to have available MHD data products generated regularly at the JSOC data center.

Various MHD models (\eegg \opencite{Suess95}; \opencite{Mikic99}; \opencite{Usmanov00}; \opencite{Nakamizo09}; \opencite{Feng10}; \opencite{Usmanov11}) have applied sophisticated models of coronal heating and acceleration, such as \myAlfven wave decay and the radiation process, to successfully produce realistic features of the solar corona and solar wind, such as steep radial gradients (or acceleration) of the solar wind speed in the solar corona, final velocities in distant regions, and temperatures in coronal streamers, active regions, and coronal holes.
In general, these models are associated with the heat conduction term that has a much shorter time scale than the global \myAlfven waves; thus it requires intensive computation to compute them.
Unfortunately, as one of the modules at the HMI data pipeline, it is not practical for us to implement such computationally expensive models.

The spatial resolution of the MHD simulation is another factor that we have to consider in estimating the computational resources needed.
In the spherical coordinate system, a straightforwardly constructed grid system will have a longitudinal grid size of $r \Delta\phi\sin(\delta\theta) \approx r\Delta\phi\Delta\theta$ near the polar axis, which is approximately proportional to $(\Delta\phi)^3$ if $\Delta\phi\approx\Delta\theta$.
Roughly, the time step is inversely proportional to the smallest size of the grid, and the spatially high-resolution simulation cannot be regularly conducted with limited computational resources.

In practice, low-resolution simulations generated in a timely manner will be more beneficial as the standard derivative product in the HMI data pipeline.
For example, quick-look simulation data will help us grasp the current state of the solar corona and solar wind, and a sudden change found in quick-look data offers early warnings to recognize at least the possibility of global scale variations worth further examination.

Since early 2008, well before the HMI observations began in May 2010, we started running daily MHD simulations using the Solar and Heliospheric Observatory (SOHO) / Michelson Doppler Imager (MDI) data.
Through preparation runs using our existing MHD code (\opencite{Hayashi05}), several choices had been tested.
Simulation plots can be found at \url{hmi.stanford.edu/MHD/daily_mhd.html}.
Several features are necessary for data products in the HMI pipeline, such as timeliness of production of near-real-time (NRT) data.
Computational resources are another factor that we have to take into account.
Based upon the experience gained through the test runs, we choose to run low-resolution ($\approx 5.6$ degrees) MHD simulations on a daily, NRT basis, and medium-resolution ($\approx 2,8$ degrees) simulation once each Carrington rotation, with a set of simulation settings that have been optimized to provide the optimal combination of the MHD simulations and HMI observations to the solar physics community as an HMI data product.
The MHD data as well as source codes for this module can be found at \url{jsoc.stanford.edu}.

This article describes the details of the magnetic-field map production process, the MHD model settings, and the new data products available at the JSOC.
This article is organized as follows:
In Section~\ref{SCTmagmap} we describe two types of input magnetic-field maps, the synoptic maps and the synchronic maps, regularly generated at the JSOC.
In Section~\ref{SCTmhd}, details of the MHD models employed are given.
In Section~\ref{SCTjsoc}, information needed to access and specify the MHD data using the JSOC data interface is provided.
Section~\ref{SCTdemo} demonstrates the results of MHD simulations and makes comparisons with the AIA coronal images and the near-Earth \myinsitu OMNIweb data set over a selected period from late 2010 to early 2011.
Summaries and remarks are given in Section~\ref{SCTsmmry}

\section{Synoptic and Synchronic Maps of the Magnetogram Data}
\label{SCTmagmap}
Synoptic maps of the magnetic field have been widely used for models of the global corona, such as the potential-field source-surface (PFSS) (\opencite{Schatten69}; \opencite{Altschuler69}) and MHD simulation models.
We can specify realistic conditions of the magnetic field at a time of interest for the coronal models by using the whole-Sun map of the observed solar photospheric magnetic field data in a synoptic map or a synchronic frame.

When making full-surface maps on a daily, NRT basis, we need to implement some extra procedures, such as reducing the distortion of the map at high latitudes due to the differences between the differential solar rotation and the assumed rigid rotation.
We have developed an improved whole-Sun map, called the synchronic frame (\eegg \opencite{Zhao99}), and we use it to create the daily updated maps in the HMI data pipeline.

\subsection{The Synoptic Chart and Synchronic Frame in JSOC}
\begin{figure}
\centerline{
 \includegraphics[width=0.8\textwidth,clip=]{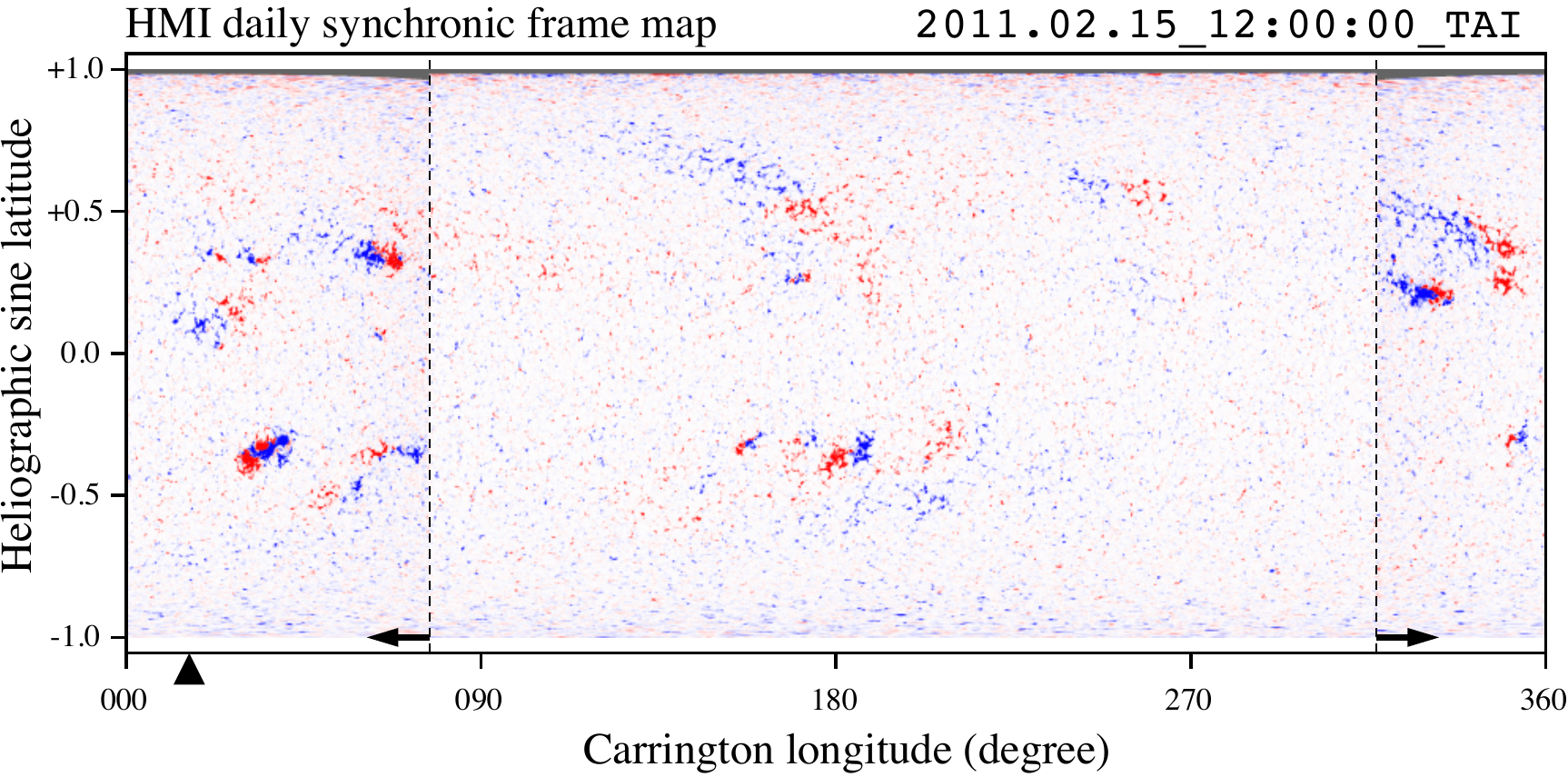}
}
\vspace*{1truecm}
\centerline{
 \includegraphics[width=0.8\textwidth,clip=]{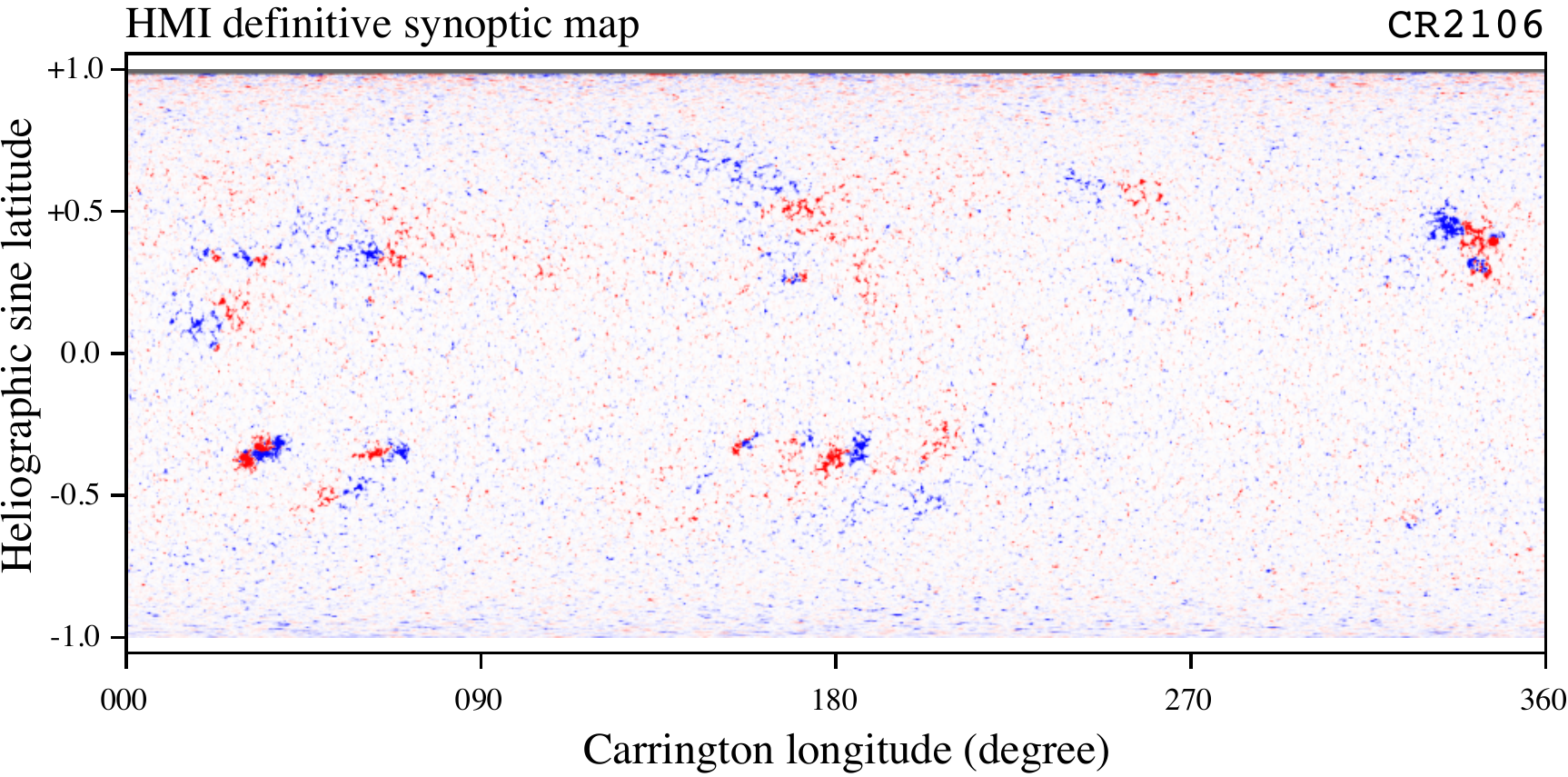}
}
\caption{%
The synchronic frame for 15 February 2011 (top) and synoptic map for CR 2106, from 20 January 2011 to 16 February (bottom).
In the top panel, two dotted vertical lines are placed at the bounding longitudes of the updated part.
Two arrows indicate the 120-degree updated region, and a triangle is placed at about 16 degrees of the Carrington longitude that corresponds to the central meridian on 15 February 2011, 12:00 UT.
Blue (red) represents the positive (negative) polarity.
The colors are truncated at $\pm$ 100 Gauss.
}%
\label{fig1}
\end{figure}

Figure \ref{fig1} shows an example of the once-per-CR standard synoptic map and daily synchronic map of $B_r$ that the MHD modules use.

In the HMI data pipeline, a standard Carrington synoptic chart is made as follows.
First, the line-of-sight field is converted to the radial component by dividing by cosine of the center-to-limb angle (\opencite{Wang92}).
The conversion to the radial-field component means that the B-zero angle ($B_\mysun$) has been accounted for, and no additional correction for the dependence on $B_\mysun$ and other projection effects is required.
The radial component is then remapped to a Carrington coordinate grid.
Each point in the grid has been adjusted to the time of its central meridian in order to minimize additional smearing due to the differential rotation, as suggested by \inlinecite{Ulrich02}.
The remapped grid retains the HMI's spatial resolution at the disk center, i.e. 0.03 degrees.
It is then reduced to 0.1 degrees by convolving with a two-dimensional Gaussian with a half width of three pixels.
The field strength is then averaged from all of the contributing remapped magnetograms.
Currently, the average is done from twenty 720-second magnetograms that contribute pixels observed around the central meridian.
Therefore, the effective temporal width of the HMI synoptic map is about four hours at each Carrington longitude, i.e. within two hours of central meridian passage, or within about 1.2 degrees of longitude.
The line-of-sight field map is derived from the radial field chart by multiplying the cosine of the latitude for each grid pixel.
Therefore, the data are Equator-centered, meridian-centered, line-of-sight field values as if observed from a point on the solar Equator.

A daily updated synchronic frame radial field is produced from the remapped HMI 720-second radial-field magnetograms and the standard Carrington synoptic map of radial field: the remapped magnetograms make up a 120-degree daily updated region between the two edges at 60 degrees of longitude from the central meridian at UT noon.
A synoptic chart makes up the rest of the daily updated synchronic map.
The data values in the daily updated part are four-hour averages of the remapped magnetograms.
This four-hour averaging, using twenty 720-second magnetograms, assures noise in the updated part to match that of the synoptic chart.
There are two types of daily updated synoptic maps.
One is from NRT 720-second line-of-sight magnetograms.
The other is from the definitive 720-second line-of-sight magnetograms generated with about 45-day delay.
The size of map is 3600 by 1440 with longitude in $x$-axis and $\sin$(latitude) in $y$-axis.
Smaller size synoptic maps (720 $\times$ 360) are also produced.

Table \ref{tbl1} tabulates the series names, the topmost identifier in the JSOC database, for the standard synoptic map and daily updated synchronic frames.
The MHD modules use the radial component maps (marked with asterisks (*) in Table \ref{tbl1}).
As soon as a NRT synchronic frame for UT noon is generated, a low-resolution NRT MHD run starts, and the MHD data becomes available around 22 UT at latest.
The definitive version of the daily map and the standard once-per-CR synoptic map are usually generated 45 days after the last observation, when all necessary calibration processes are completed.

\section{MHD Model of the Global Corona}\label{SCTmhd}%
The main part of the three-dimensional time-dependent MHD simulation code is almost identical to the one used in our previous works (\iiee \opencite{Hayashi05}; \opencite{Hayashi08}, and \opencite{Hayashi13}).
In brief, it employs fairly standard concepts in time-dependent MHD simulation, such as the Total Variation Diminishing (TVD) (\eegg \opencite{Brio88}; \opencite{Roe96}) and Monotonic Upstream Scheme for Conservation Laws (MUSCL) (\eegg \opencite{vanLeer79}).
One notable feature in our code is that the concept of the projected normal characteristics method (\opencite{Nakagawa87}; \opencite{Wu87}; \opencite{Han88}) is applied to treat the solar-surface inner boundary sphere.
This method can provide a physically consistent environment without artifacts of incoming waves from outside the simulated volume, keeping the sinusoidality of the magnetic field (\eegg \opencite{Yeh85}).
The details of the model are given by \inlinecite{Hayashi05} and references therein.

To speed up the simulation, we employ the Lax-Friedrichs method instead of the linearized Riemann solver that contains computationally expensive operations of eight-by-eight left and right eigen-vector matrices of MHD hyperbolic equation system.
We note that because the concept of the normal projected characteristic method is applied, the same eight-by-eight eigen-matrices are still used in this simplified version.

\subsection{MHD Model}
Instead of using a coronal heating and acceleration model, we simply assume a near-isothermal polytropic gas with specific heat ratio $[\gamma]$ of 1.05 to make the trans-\myAlfvenic solar wind.
From provisional tests with various specific heat ratios from 1.001 to 1.2, we found that 1.05 is best because the ratio can create a moderate contrast in coronal density and plasma speed.

It is easily expected that the polytrope model may result in higher solar wind speed near the heliospheric current sheet (HCS) originating from the slow-speed region just above the closed-field coronal streamer where higher density is always associated with higher temperature in the modeled solar wind.
In addition, the contrast of the simulated solar wind speed at 1 AU is generally very small.
Thus, we have to be careful when using the data at 1 AU, though this does not mean that generality in the nonlinear MHD interaction processes in the sub/trans-\myAlfvenic solar corona had been lost.
Because the near-isothermal polytrope model has been a standard reference in solar-wind studies probably since the pioneering work of \inlinecite{Parker58}, it is reasonable to use this model even with the caveats for the distant, 1 AU region.
Overall, global coronal magnetic-field structures are not sensitive to the heating and acceleration processes, but we still need to be cautious.
For example, one of the coronal features that might be sensitive to the choice of the coronal heating is the unipolar boundary layer (UBL) in interplanetary space (\opencite{Zhao03}) or pseudo-streamers in the corona, below which two compact closed-field structures are located between two coronal holes of the same magnetic polarity.
In this configuration, the pressure balance between the magnetic field and plasma might be a critical factor determining whether the simulated system will have UBL or two small closed-field regions with a narrow coronal hole in between.

With the assumption of the near-isothermal polytrope, the basic equations to be solved are the time-dependent MHD equations in the frame rotating at the sidereal angular velocity of solar rotation, $\vec{\Omega}$;
\begin{eqnarray}
   \frac{\partial\varrho}{\partial t}
&=&
   -\nabla\cdot(\varrho\vec{V})
\label{mhda}\\
   \frac{\partial(\varrho\vec{V})}{\partial t}
&=&
   -\nabla\cdot
   \left(
      P_g +\varrho\,\vec{V}:\vec{V}
     -\frac{1}{4\pi}\vec{B}:\vec{B}+\frac{B^2}{8\pi}
   \right)
\nonumber \\
&+&\varrho\,
   \bigl[
      \vec{g}
     +(\vec{\Omega}\times\vec{r})\times\vec{\Omega}
     +2\,\vec{V}\times\vec{\Omega}
   \bigr]
  -(\nabla\cdot\vec{B})\vec{B} 
\label{mhdb}\\
  \frac{\partial\vec{B}}{\partial t}
&=&
  -\nabla\cdot(\vec{V}:\vec{B}-\vec{B}:\vec{V}) 
  -(\nabla\cdot\vec{B})\vec{V} 
\label{mhdc}
\end{eqnarray}%
and
\begin{eqnarray}%
  \frac{\partial{\cal E}}{\partial t}
&=&
  -\nabla\cdot
   \left(
     {\cal E}+P_g-\frac{1}{4\pi}(\vec{V}\times\vec{B})\times\vec{B}
   \right)
\nonumber\\
&+& \varrho\,\vec{V}\cdot
   \left(
     \vec{g}+(\vec{\Omega}\times\vec{r})\times\vec{\Omega}
   \right)
  -(\nabla\cdot\vec{B})(\vec{V}\cdot\vec{B}),
\label{mhdd}
\end{eqnarray}%
where $\varrho, \vec{V}, \vec{B}$, $P_g$, ${\cal E}$, $\vec{r}$, $t$, $\vec{g}$ and $\gamma$ are mass density, velocity of plasma flow viewed in the frame rotating with the angular velocity $\vec{\Omega}=2\pi/25.3$ radian day${}^{-1}$ (or 14.2 degrees day${}^{-1}$), magnetic-field vector, gas pressure, energy density ${\cal E}=\varrho v^2/2+P_g/(\gamma-1)+B^2/2$, position vector originating at the center of the Sun, time, solar gravitational force $\vec{g}=-GM_\mysun/r^3\cdot\vec{r}$ and specific heat ratio, respectively.
In this notation, `$:$' expresses the dyadic tensor product of two vectors.
The specific heat ratio, $[\gamma=1.05]$ is constant everywhere.

The normalizing factors (or typical values) of density, temperature, flow speed and magnetic field are set as, $n_\mysubzero=9.0641\times 10^4$ count \mypcc, $T_\mysubzero = 1\times 10^6$ \mykelvin, $V_\mysubzero=131.66$ \mykmps, and $B_\mysubzero=1.817\times 10^{-2}$ Gauss, respectively.
Table \ref{tbl2} tabulates setting of parameters used in this model.

The initial values of plasma quantities are set as the Parker solution, with the velocity and temperature equal to the normalizing factors at its critical point (at $r=3.8282\times 10^6$ km or 5.47 solar radii).
The typical density is selected such that the density on the bottom boundary sphere, $n_\mysubb$, will be $2\times 10^8$(count \mypcc).
The normalization factor of the magnetic field is chosen such that the typical magnetic pressure is equal to the typical ram pressure, $B_\mysubzero^2/8\pi = n_\mysubzero V_\mysubzero^2/2$.
The typical thermal pressure of perfectly ionized hydrogen gas $P_\mysubzero$ is equal to $2n_\mysubzero k_\mysubB T_\mysubzero$ where $k_\mysubB$ is the Boltzmann constant.
The initial magnetic field is the solution of the PFSS model with the boundary map given from the upstream modules in the HMI data pipeline.

Because of the cell-centered grid system in our model, the divergence of magnetic field will inevitably differ from zero.
To address this problem, the terms associated with the divergence of the magnetic field ($\nabla\cdot\vec{B}$) are added to the right hand side of Equations (\ref{mhdb}), (\ref{mhdc}), and (\ref{mhdd}) so that the effect of the numerical monopole will be minimized and so that the numerical magnetic monopole will move with the plasma material and finally be swept out of the computational domain (\opencite{Brackbill80}; \opencite{Powell94}; \opencite{Powell99}; \opencite{Toth00}).

\begin{figure}
\centerline{
 \includegraphics[width=0.35\textwidth,clip=]{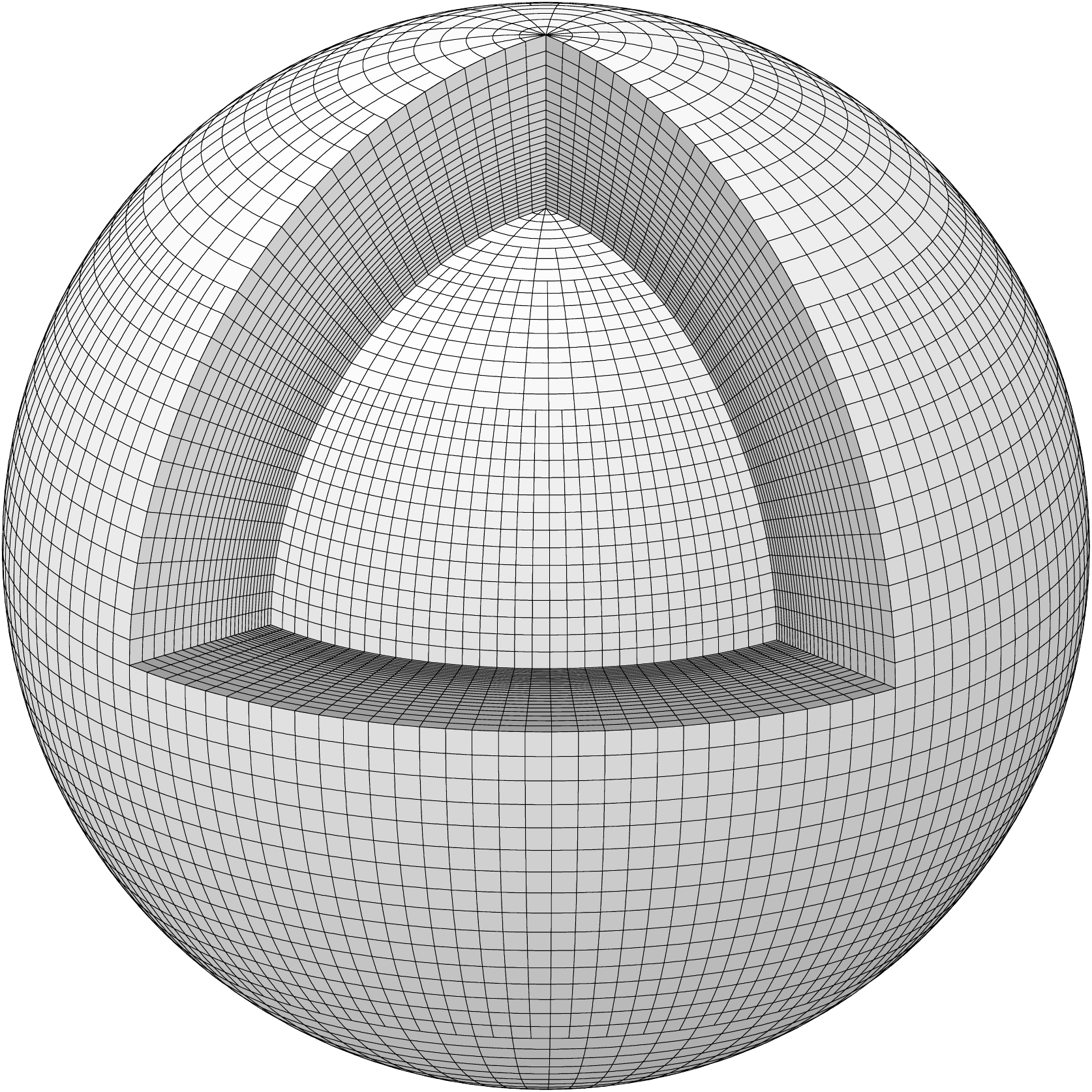}
 \hspace*{1truecm}
 \includegraphics[width=0.35\textwidth,clip=]{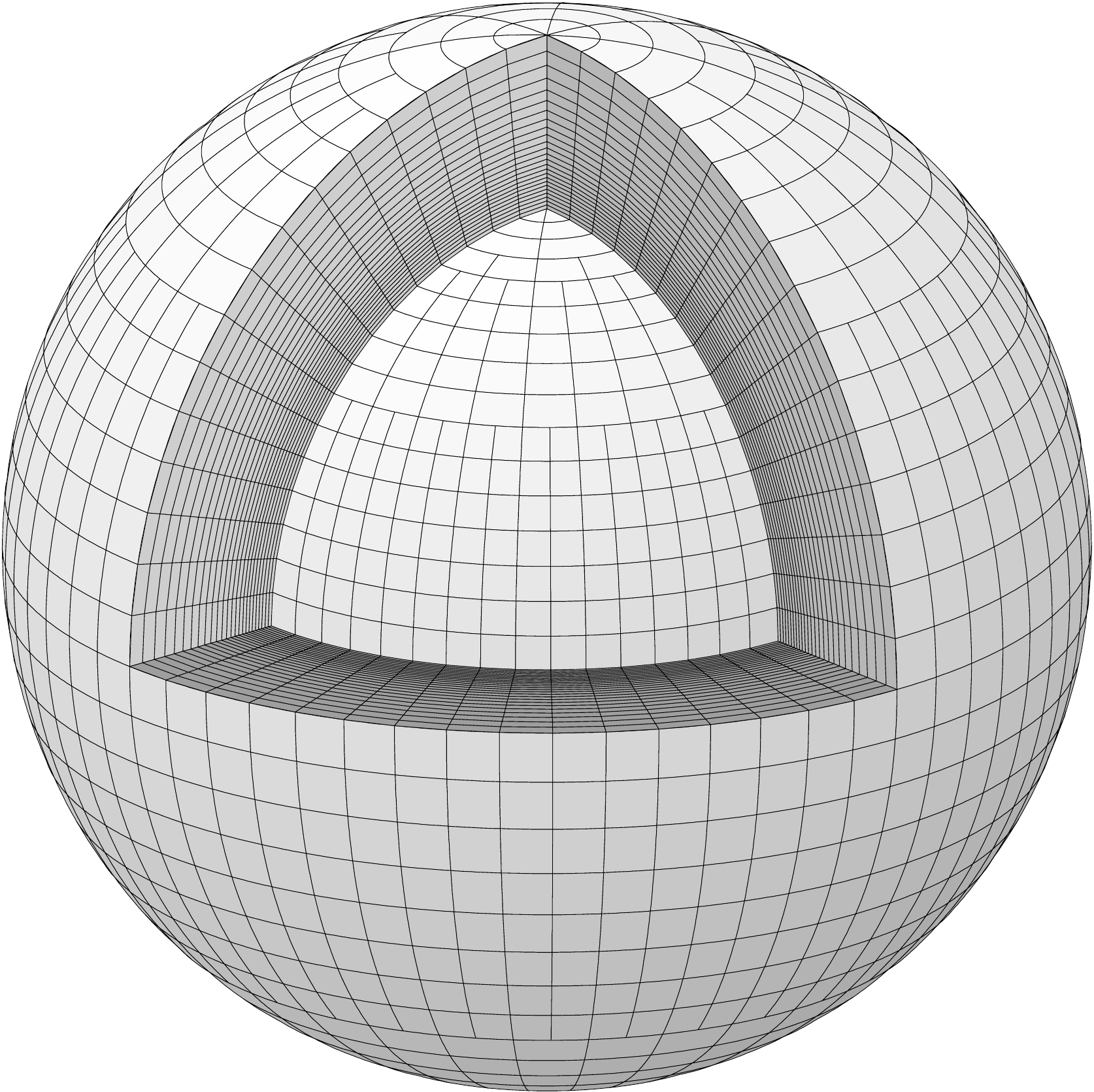}
}
\caption{%
Grid systems of the simulation.
(left) the grid system of moderate resolution that is used for definitive once-per-CR run.
(right) the lower-resolution grid system.
For visibility, only the inner 25 of 72 spherical layers in the radial direction are drawn.
The finite volume method (FVM) merges the cells near the pole and uses appropriately averaged values.
This averaging mitigates the CFL constraint and helps speed up the calculation while preserving the conservative quantities such as mass and total energy.
}%
\label{fig2}
\end{figure}

The grid system is constructed in a spherical coordinate system so that the input whole-Sun map can be hosted straightforwardly.
To discretize the space, we employ the concept of the finite volume method (FVM) (\opencite{Tanaka95}).
From our provisional tests, we set the simulated space to be from 1.01 Rs to 50 Rs in the radial direction, and from 0 to $\pi$ in the latitudinal direction.
This range of heliocentric distance is divided into 72 grid cells (excluding the ghost cells).
The innermost five and outermost five grids have uniform radial grid sizes.
In the simulation, the latitude runs 180 degrees from North to South (co-latitude) and the longitude runs from West to East, with equi-angle grid system.
The numbers of grids in the latitudinal and longitudinal directions are 32 and 64 in the low-resolution daily simulation, and 64 and 128 in the medium-resolution once-per-CR run.
If we straightforwardly apply the equi-angle grid system, the size of cells near the poles are small and the time step $[\Delta t]$ will be problematically small.
To mitigate the computational cost due to the severe CFL condition, numerical cells near the poles are merged: as in \inlinecite{Hayashi08}, the number of cells to be longitudinally merged is calculated to be $N$th power of 2, $2^N$, with $N$ being the smallest integer satisfying
\begin{equation}
 2^N(r\Delta\phi\sin\theta)\ge\min({\Delta}r,r\Delta\theta)
\end{equation}
at each latitude $[\theta]$.
Figure \ref{fig2} shows the grid systems used in our coronal MHD model.

\subsection{Three Simulation Settings}\label{SCTsetting}
\begin{figure}
\centerline{
 \includegraphics[width=1.0\textwidth,clip=]{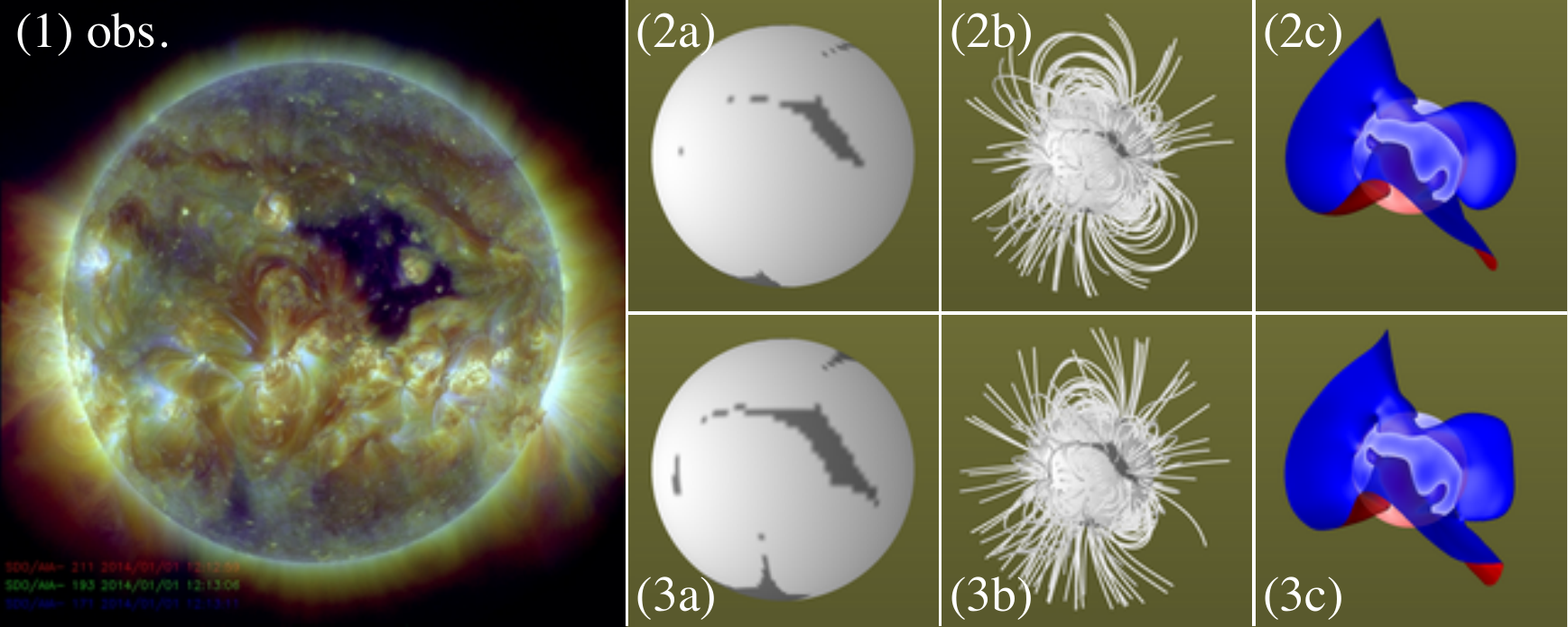}
}
\caption{%
An example of the simulation results and comparison with the AIA measurement.
Frame 1 is the three-color composite of AIA images taken around noon UT, 1 January 2014.
Frames 2a to 2c show the base of the open-field region, coronal field lines, and the contour surface of the zero value of the radial component of the initial PFSS magnetic field.
Frames 3a to 3c show the same properties as Frames 2a to 2c except the time-relaxed MHD solution is used.
The magnetic-field map given to the initial PFSS and MHD model is for CR 2145, whose central time is approximately the date of the AIA observation.
The viewpoint in Frames 2a to 3c is set at 182.83 degrees Carrington longitude and 3.2 degrees South heliographic latitude, the approximate position of the Earth at the time of the AIA observations.
}%
\label{fig3}
\end{figure}

Considering various factors, such as required computational resources and timeliness of data delivery, we select three types of MHD simulation: 
(A) Medium-resolution simulation (with 72, 64, and 128 grid points for the radial, latitudinal and longitudinal directions, respectively) that is computed about once in four weeks (once-per-CR). This choice uses a Gaussian-smoothed definitive synoptic map. The initial potential field is calculated with the iterative Laplace solver so that the input map can be identical to the smoothed map.
(B) Low-resolution simulation (with 72, 32 and 64 grid points for the three directions) using the definitive daily updated map.
The fifth-order spherical harmonic polynomials are used to calculate the initial and boundary magnetic field.
(C) Low-resolution simulation with the same setting as (B) except that this choice uses the near-real-time version of the input magnetic field instead of the definitive one.
Figure \ref{fig3} shows the field lines and current sheet in the corona obtained with a medium-resolution simulation (A) using a definitive map for CR 2145, as an example. 

The three types of the simulation use identical code except for the grid size.
Each run simulates a 40-hour time-relaxation process starting from the initial PFSS and Parker solution.
In the current JSOC cluster computer system, the inter-node MPI job capability is disabled, in order to optimize the system for processing a large number of single-CPU jobs.
Therefore, the number of CPU cores available to one particular parallel computation is limited to eight.
On an 8-CPU node, the efficiency of our MHD code parallelized with OpenMP is almost 800$\,\%$, and it takes about three hours to complete a low-resolution daily run and about one day to complete the once-per-CR medium-resolution simulation.

The first and second simulation runs start after the calibration of the full-disk magnetograms is completed and the definitive version of the whole-Sun maps have been created, which usually takes place a month after the observation.
The last simulation (C) starts immediately after the daily NRT synchronic map is generated, usually around 14 UT, and the simulated data can be available around 17 UT.
The time lag of about five hours between the last full-disk magnetogram observation and the simulation data production may be reasonably acceptable; because the lag is much shorter than the typical travel time of the solar wind from Sun to the Earth, two to four days.

\section{Output Data at JSOC}\label{SCTjsoc}%
The MHD data are available, along with the other HMI products, at the website, \url{jsoc.stanford.edu/ajax/lookdata.html}.
In most cases, users have to provide three pieces of information to specify particular data: the series name, the primary keyword value, and the segment.
The series name is an identifier, at the topmost of the database-tree structure, to specify the type of data.
The primary keyword is to specify a record (a minimum set of data to be physically or observationally meaningful), typically a time.
The segment, the lowermost layer of the database, specifies the name of a physical or observational variable.

\subsection{Series Name, Record, Name of Segment, and Keyword}
Table \ref{tbl3} tabulates the series names of the MHD products, (A), (B), and (C), together with the input magnetic-field map.

The primary keyword, an index of time, of the definitive once-per-CR simulation data is the Carrington Rotation number.
For example, the MHD data for CR 2140 is specified as \texttt{hmi.MHDcorona[2140]}.

As for the daily-MHD data, each data record is specified with the time \texttt{T\_REC} of the magnetic-field data map that is usually denoted in TAI time format, \texttt{YYYY.MM.DD\_HH.MM.SS}.
On most days, the daily updated magnetic-field maps are generated and time-stamped around noon UT.
Thus, the hour and minute of the daily input magnetic-field and the MHD product, \texttt{HH.MM}, are \texttt{12:00}, or neighboring 12-minute slots, \texttt{11:48} or \texttt{12:12} less frequently.
The two-digit second, \texttt{[SS]}, is always zero.

The names of segments in the MHD product are notation of physical variables, i.e., \textsf{N}, \textsf{T}, \textsf{Vr}, \textsf{Vt}, \textsf{Vp}, \textsf{Br}, \textsf{Bt}, and \textsf{Bp} (plasma number density, temperature, the radial, latitudinal, and longitudinal component of plasma velocity, and the three components of the magnetic field, respectively).
Table \ref{tbl4} lists the segments, the physical variables, and the units used in storing the data values.

\subsection{Keywords}
In the JSOC database context, a large set of keywords collects information relating to the observation, calibration, or other post-processes, such as the time of observation, telescope setting and state of the spacecraft, and code version number.
Because the FITS file generated at the JSOC website contains most of the JSOC keywords in the header field, it is in most cases appropriate to regard the JSOC keywords as the FITS header information.

Many keywords of the MHD products are copied from the input magnetic-field data set, and some of them are same values and others may be updated accordingly.
There are several keywords newly added in order to describe basic nature of the MHD simulation.
Tables \ref{tbl5} and \ref{tbl6} summarize the most relevant keywords, those succeeded from the input magnetic-field data, and those updated by the MHD module, respectively.
A complete list of keywords can be found at the JSOC webpage, \url{jsoc.stanford.edu/ajax/lookdata.html}.

\subsection{Data Grid System}
Our MHD model employs a non-uniform grid size along the radius.
For various reasons, the number of grid points in latitude and longitude have to be a power of two ($2^5$ for the low-resolution and $2^6$ for the medium-resolution simulation).
This grid size was chosen after taking into account the steep gradient of the density and temperature near the solar surface and some computational efficiency.
This simulation grid system may not be convenient for general use, so we reorganize the data.

The MHD data cubes available in the JSOC database segments are linear-interpolated spatially into equi-distance for the radial direction and equi-angle for the latitudinal and longitudinal directions.
We choose a right-hand system where the first position address runs for the longitudinal direction (from East to West in the Carrington longitude), the second does for the latitude (from South to North), and the third one does for the radius.
In this format, a three-dimensional address $[i,j,k]$ positions a grid point $[\phi_\mysubi,\theta_\mysubj,r_\mysubk]$,
\begin{equation} 
\phi_\mysubi  =    (i+1/2)\Delta\phi, \quad
\theta_\mysubj=-90+(j+1/2)\Delta\theta, \quad
r_\mysubk     =1.0+(k+1/2)\Delta r.
\end{equation}
Notice that here each address index starts from zero.

We note that this data coordinate system has an opposite sense in latitude to those of the MHD code where the (co-)latitude directs from North to South.
To be fully self-consistent in the production data, the segments, \textsf{Vt} (for $V_\theta$) and \textsf{Bt} (for $B_\theta$), are set in a way such that positive values stand for the component directing toward the North.

The longitude of the input synoptic/synchronic maps in the JSOC database is defined as the Carrington time, which runs from West to East, in an opposite sense to the ordinary definition of longitude.
As the downstream product of the magnetic-field map, the MHD production data have to use the same definition.
In order to keep the right-hand system, on the other hand, we want to keep the address for the longitudinal direction in the cubic data.
To satisfy all requirements, we define the integer address running from East to West in the same manner as the ordinary longitude in spherical coordinates, while its incremental step is defined to be a negative value.
In this definition, the Carrington time can be calculated as
\begin{equation}
  \phi_{{\rm CR},i}  = 360 + (i+1/2)\Delta\phi_{\rm CR}.
\end{equation}
The angular increment steps in the longitudinal and latitudinal directions $[\Delta\phi_{\rm CR}$ and $\Delta\theta]$ are recorded as a value of the keywords \texttt{CDELT1} and \texttt{CDELT2}, respectively.
As summarized in Table \ref{tbl5}, the values are fixed in the same series, and the values are multiple of 2.5 degrees, \iiee $(\Delta\phi_{\rm CR},\Delta\theta)$ are $(-2.5, 2.5)$ for the medium-resolution simulation (A), and $(-5.0,5.0)$ for the daily low-resolution simulations (B) and (C).

\begin{figure}
\centerline{
 \includegraphics[width=0.6\textwidth,clip=]{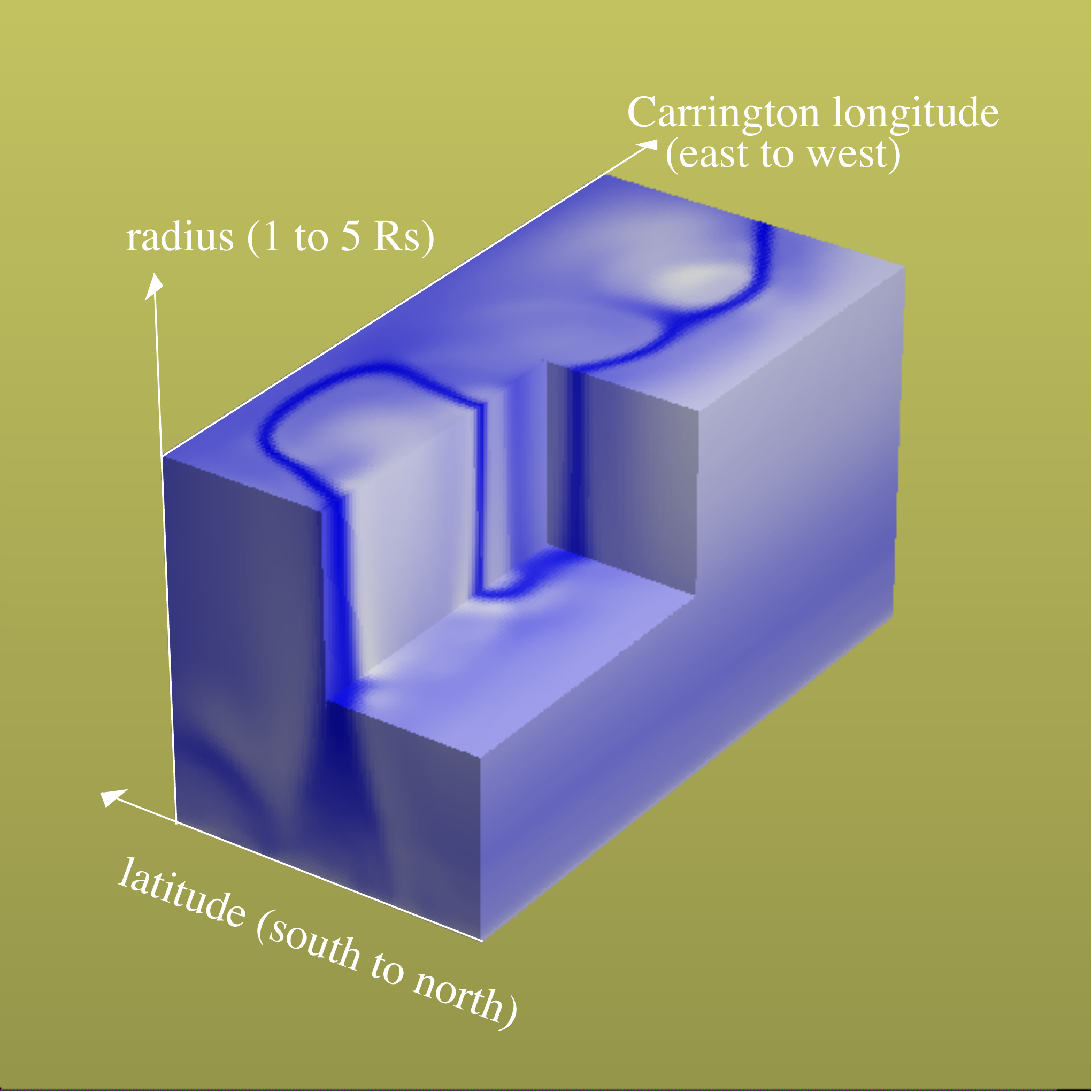}
}
\caption{%
A cut-away view of cubic, three-dimensional synoptic data.
The plasma density $[N]$ normalized with average value at each height (heliocentric distance) obtained with simulation (A) for CR 2145 is shown.
On the top face (at 5.0$R_\mysun$), the high-density streamer belt is visible.
On the left face (at zero degrees longitude), the cross-section of a streamer is visible near the heliographic Equator, and another high-density region corresponding to the polar crown closed-field region can be seen at northern high latitude.
}%
\label{fig4}
\end{figure}

Figure \ref{fig4} shows a cut-away view of the coronal density normalized with its average at each height, as an example for CR 2145.
For this period, a highest-density belt is seen on the topmost layer at 5.0 solar radii, which corresponds to the base of the HCS and the topmost part of coronal streamers.
The evolution of the streamer with respect to height can be seen on the left face (the meridional plane at longitude 0 degree).
Below about three solar radii, the field lines are closed, and the outermost layer of the closed-field structure had higher density than the outer part of a coronal hole and the inner part of the streamer.
Above about three solar radii, many of the magnetic fields are open to the interplanetary space, and the plasma is no longer confined.
Still, the speed is slower and the density is higher than in the surroundings, and the high-density regions can reach the interplanetary space.

On the left, another short-height closed-field region can be seen in the northern hemisphere (left). 
This high-density structure near the pole corresponds to the surroundings of the polar crown.
The structure forms a dome over the (northern) pole whose magnetic polarity is opposite to that of the surrounding high latitudes.
This structure may be somewhat common in the maximum phase of solar activity; the reversal of the interplanetary polar field takes place before the highest-latitude solar-surface magnetic field reverses its polarity.

\section{Data Demonstrations}\label{SCTdemo}%
The MHD modules provide the plasma variables (density and temperature) as well as the magnetic field in the solar corona.
Because the derived structures are a consequence of the non-linear interaction between the plasma and magnetic field, the derived magnetic-field structures are, specifically at the upper corona, are different from those obtained through the PFSS model.
This section exhibits the coronal plasma density, shapes of open-field region, and the magnetic field escaping to interplanetary space, derived from the daily and once-per-CR simulations.
We chose a period from late 2010 to early 2011.

\subsection{Coronal Density}
First, we examine the plasma density in the corona, which is one of the derivative products from the photospheric HMI observables.
The derived density is generally highest at the base of the HCS near the top of coronal streamers, and noticeably higher at the unipolar boundary layers than in the open-field corona holes.
Therefore, the computed density is a good proxy of the global magnetic field structure.

\begin{figure}
\centerline{
 \includegraphics[width=1.0\textwidth,clip=]{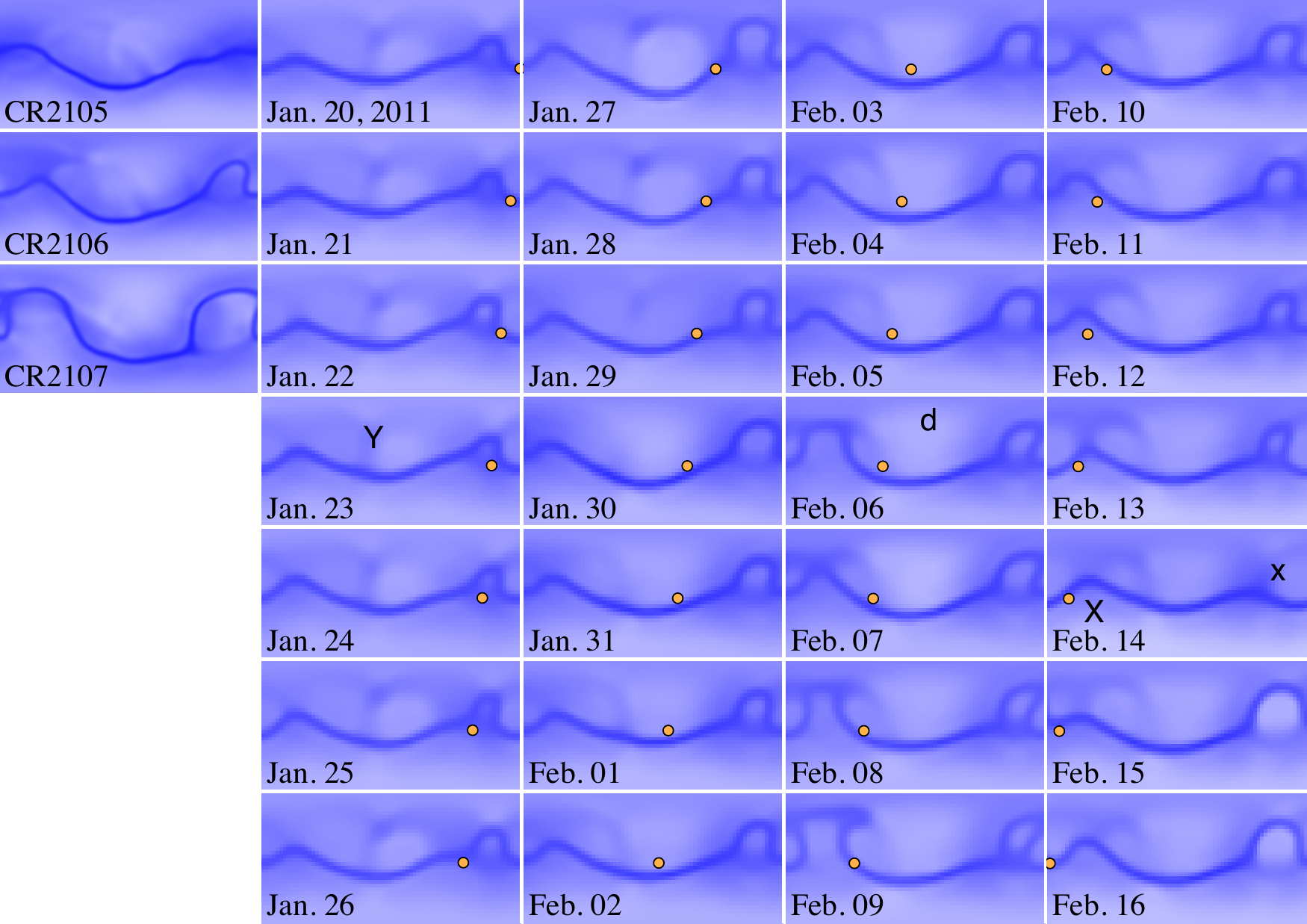}
}
\caption{%
Density in the topmost layer of the data cube at about 5.0 $R_\mysun$.
The white-blue color table spans density range from $0.9\times 10^5$ (white) to $1.4 \times 10^5$ count \mypcc\, (blue).
The highest density (dark blue) corresponds to the lower-speed current sheet or the uppermost part of a stagnant coronal streamer, and the brighter blue corresponds to open-field regions with lower density and faster outward flow.
In the leftmost column, the density maps derived from the medium-resolution simulation (A), for three consecutive Carrington Rotation numbers from 2105 to 2107, are shown for reference.
The second to fifth column show the density maps obtained through the low-resolution daily simulation using definitive daily maps (B), from 2011 Jan 20 (corresponding the start date of CR 2106) to 16 February 2011 (the last day of CR 2106).
An orange-colored circle is placed at the approximate position of the Earth.
The mark `Y' on 23 January map indicates where three unipolar boundary layers are merging, and the `d' on 6 February indicates a disappearance of the unipolar boundary layer.
The mark `x' indicates approximate position of the two Active Regions, NOAA 11147 and 11161, which appeared at mostly the same heliographic position in two consecutive solar rotation periods.
The mark `X' on 14 February is placed at the position of the NOAA 11158.
}
\label{fig5}
\end{figure}

Figure \ref{fig5} shows the density structures at 5 ${\rm R}_\mysun$, the topmost data layer in the production data cube, over 28 days from 20 January to 16 February, derived from the daily simulation (B).
The same-format plots of the higher-resolution once-per-CR simulation (A) from 2105 to 2107 are shown in the leftmost column (the central time of CR 2106 falls on 3 February; the approximate center of the 28-day time span), for reference.
Overall, the highest-density belt obtained from the daily simulations follow those from the once-per-CR simulation, evolving gradually on the time scale of a few weeks to a month.
At the same time, however, we also find some daily variations can be sudden and take place in a day or two.
The merit of the daily simulation is that it can catch such sudden changes in the global corona that might be missed when we conduct once-per-CR simulation only.

We identify two factors that can cause sudden changes in the daily simulation results:
The first is the change in unbalanced flux in the global map, and the other is due to local, compact sudden changes such as emergence of magnetic flux forming solar active regions.
In the current version of the MHD module, we simply offset the apparent unbalanced flux in a synoptic frame by subtracting the total surface integration of $B_r$ divided by the area $[4\pi R_\mysun^2]$ evenly everywhere on the map so that the resultant map will have fully balanced magnetic field.
We recognize the possibility that this offsetting can result in changes of the position of the magnetic inversion line (MIL) on the global scale and thus the global structure of the simulated solar corona.
Such offsets are often due to active regions rotating into or out of the synoptic frame.
However, the effects of the offsetting may appear on the global scale in the simulation results, such as by displacement of the position of the high-density belt in a specific direction.
On the other hand, most of the local variations in the results from the PFSS and MHD can be attributed to local, sudden changes of the solar photospheric magnetic field, because the sensitivity of the solution of the PFSS (or Laplace equation) decreases very quickly with distance, thus, the solutions of the MHD simulation using the PFSS solution for the initial value have the same tendency.

We can find belts with moderately high density (i.e. lower than those at the base of HCS but higher than the surroundings ordinary coronal holes), in the north of the HCS highest-density belt.
These regions correspond to unipolar boundary layers (UBL) or pseudo-streamers where the plasma and magnetic flux from two coronal holes with the same polarity interact to form a region of moderately high density and slow plasma flow.
In the series of daily plots in Figure \ref{fig5}, a Y-shaped UBL is found in the northern hemisphere on the first several days (marked `Y' in the 23 January box), then, around 30 January, the western part of this Y-shaped structure disappeared (marked with `d' in the 6 February box).
The UBL is a consequence of a balance between the plasma and magnetic field: If the magnetic field is weaker, then two closed-field streamers with a coronal hole between them will be formed.
We need to be careful to evaluate whether simulated structures are fully matching the real ones, though we here emphasize that the changes found in the daily simulation can signal possibility of large-scale reconfiguration of coronal magnetic field.
 
In the rightmost column, we can see noticeable changes in a northward bulge of the high-density belt (marked with `x' in one box).
The bulge at about 300 degrees Carrington longitude gradually expanded and then retained its size until 14 February when the bulge structure mostly disappeared, and then it regained its size on the next day.
The daily updated maps started to include the X-flare Active Region 11158 on February 13 (marked with `X' in one box) and no significant change are found in Figure \ref{fig5}, while two other Active Regions, NOAA 11147 (seen at about 340 degrees Carrington longitude and 20 degrees North heliographic latitude in the previous solar rotation period) and NOAA 11161 (mostly at the same heliographic location; about 330 degrees of longitude and 15 degrees North), were rotating into the longitudinal range of the daily updated maps on February 15.
Therefore, the evolution of the simulated bulge structure around February 15 is mainly due to the magnetic field at the active regions with a large separation angle (about 60 degrees), NOAA 11147 and 11161 (in the northern hemisphere) and the NOAA X-flare 11158 (locating at about 30 degrees Carrington longitude and 20 degrees South).
With only once-per-CR simulations, we do not notice these possible day-by-day variations in the global solar corona and solar wind.

\subsection{Comparisons with AIA: Shape of Open-Field Coronal Holes}
A coronal hole is a good proxy of an open-field region in the solar corona.
From the simulated data, the open-field can be calculated by tracing the field line; therefore comparing simulated open field regions and dark regions in coronal observation is a good benchmark for checking the simulation data.

\begin{figure}
\centerline{
 \includegraphics[width=1.0\textwidth,clip=]{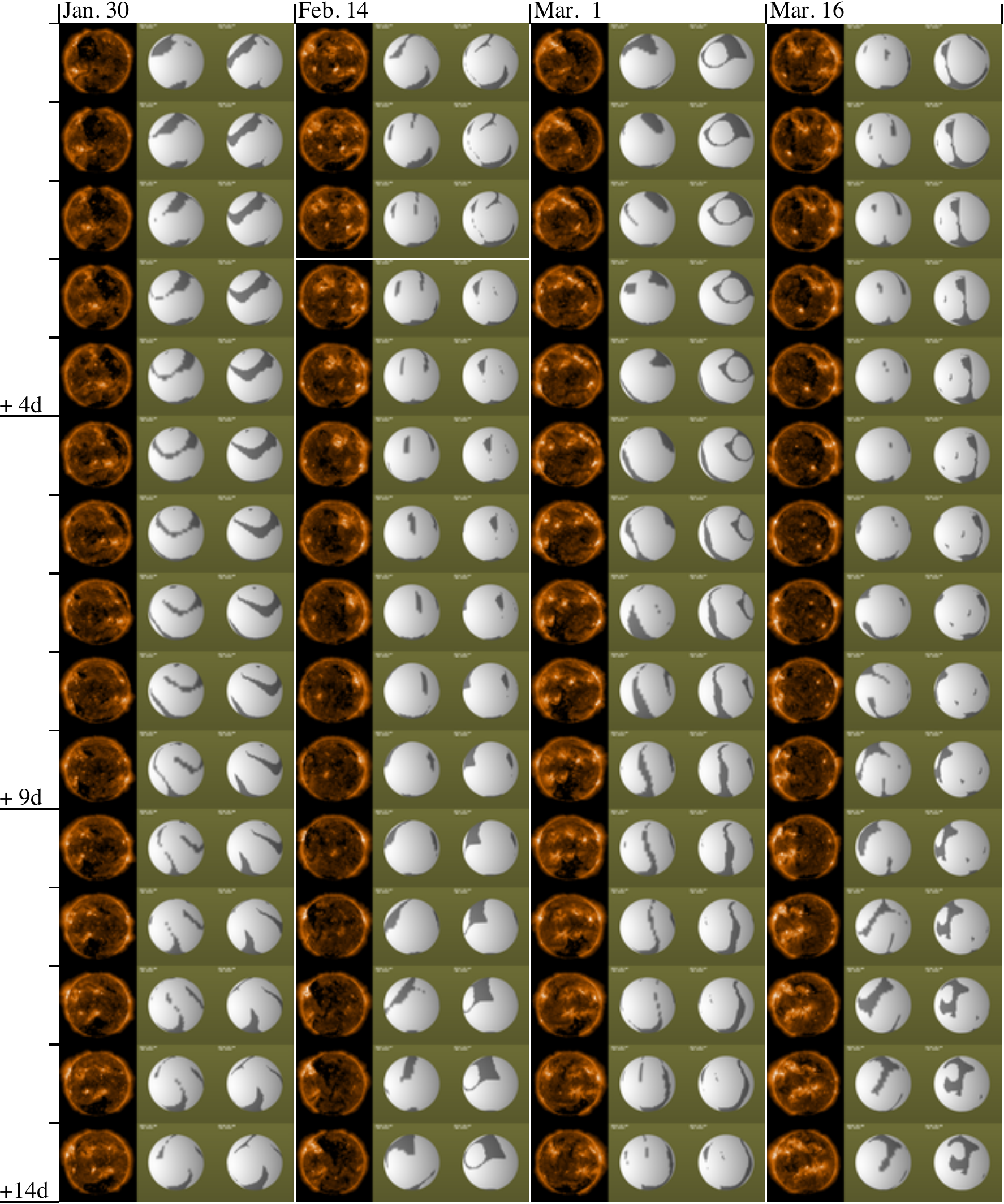}
}
\caption{%
Comparisons of shapes of the simulated open-field regions with the AIA 193\,\AA\, observation images.
In each column, from left to right, the AIA image, the daily low-resolution MHD simulation (B), and the definitive simulation results from once-per-CR simulation (A) are drawn.
The brown color for AIA image data is adjusted so that low-intensity parts will be darker and the coronal holes can be clearly visible.
The bases of the open-field (closed-field) regions of the simulated corona are colored with darker (brighter) gray.
The viewpoints in the simulation plots are set at the position of the Earth at noon each day.
Horizontal white lines are placed at the top of the box for the first day of the Carrington rotation 2107.
The Carrington rotation 2108 starts early on 16 March 2011.
}%
\label{fig6}
\end{figure}

Figure \ref{fig6} shows the AIA 193\,\AA\, image data taken around 12 UT each day, the solar-surface base of the open-field region derived from the low-resolution simulations (B) using the daily updated synchronic map, and the medium-resolution simulations (A) using the once-per-CR medium-resolution simulation, over 60 days from 30 January to 30 March 2011.
The coronal bases of the open-field region for the daily simulation (B) are viewed from the position of the Earth at the time of the daily observation.
Each plot for the simulation (A) is made by showing the simulation data for the corresponding Carrington rotation period as viewed from the Carrington longitude and heliographic latitude $[B_\mysun]$ of the Earth at Noon UT on each date.
The horizontal white lines show the dates at which the Carrington rotation number increases and thus when the input data for the simulation (A) changes.

Overall, reasonable agreements with the AIA data are found in the locations and shapes in both daily simulation (B) and the once-per-CR simulations (A), although discrepancies are also noticeable.
To achieve the best agreement, probably we need to tune further the parameters such as the plasma density and temperature and the polytropic index, although such fine-tuning is not within the scope of the simulation embedded into the HMI data pipeline.

A noticeable disagreement between the daily simulation (B) and AIA image data is found around 15 February:
the numbers of open-field regions in the Earth-facing side of the northern hemisphere simulated with the daily simulation (B) were 2, 1, and 2 on 13, 14, and 15 February, respectively, while the coronal holes in the AIA image data were rather stable and did not show such changes.
This can be attributed to the update window in the daily updated map production step, where the updated region is limited to the 60-degree window around the central meridian.
With the fixed update window size, it is possible that only part of the solar active region is included in the updated map, which can cause substantial unbalance of the magnetic-field polarity, which can affect the results of the PFSS and MHD models.

\subsection{Comparisons with Near-Earth \myinsitu Data}
The near-Earth \myinsitu data are a good reference for checking the model outputs.
For comparison with the 1-AU \myinsitu data, the radial component of the magnetic field $[B_r]$ is a suitable parameter, because this is a conservative physical quantity.
Moreover, the sign of $B_r$ in the interplanetary space is the combined consequences of the photospheric field and coronal conditions that we model.
Because the solar wind typically takes about four days to travel from the Sun to Earth, it is reasonable that the simulated value of the solar wind near the Sun at the heliographic position of the Earth predicts well the value at the Earth four days later.
In the same way, the state of the solar wind at the Earth at an instant is linked to the near-Sun value at a Carrington longitude about 53 degrees West of the current Sun--Earth direction.

\begin{figure}
\centerline{
 \includegraphics[width=1.0\textwidth,clip=]{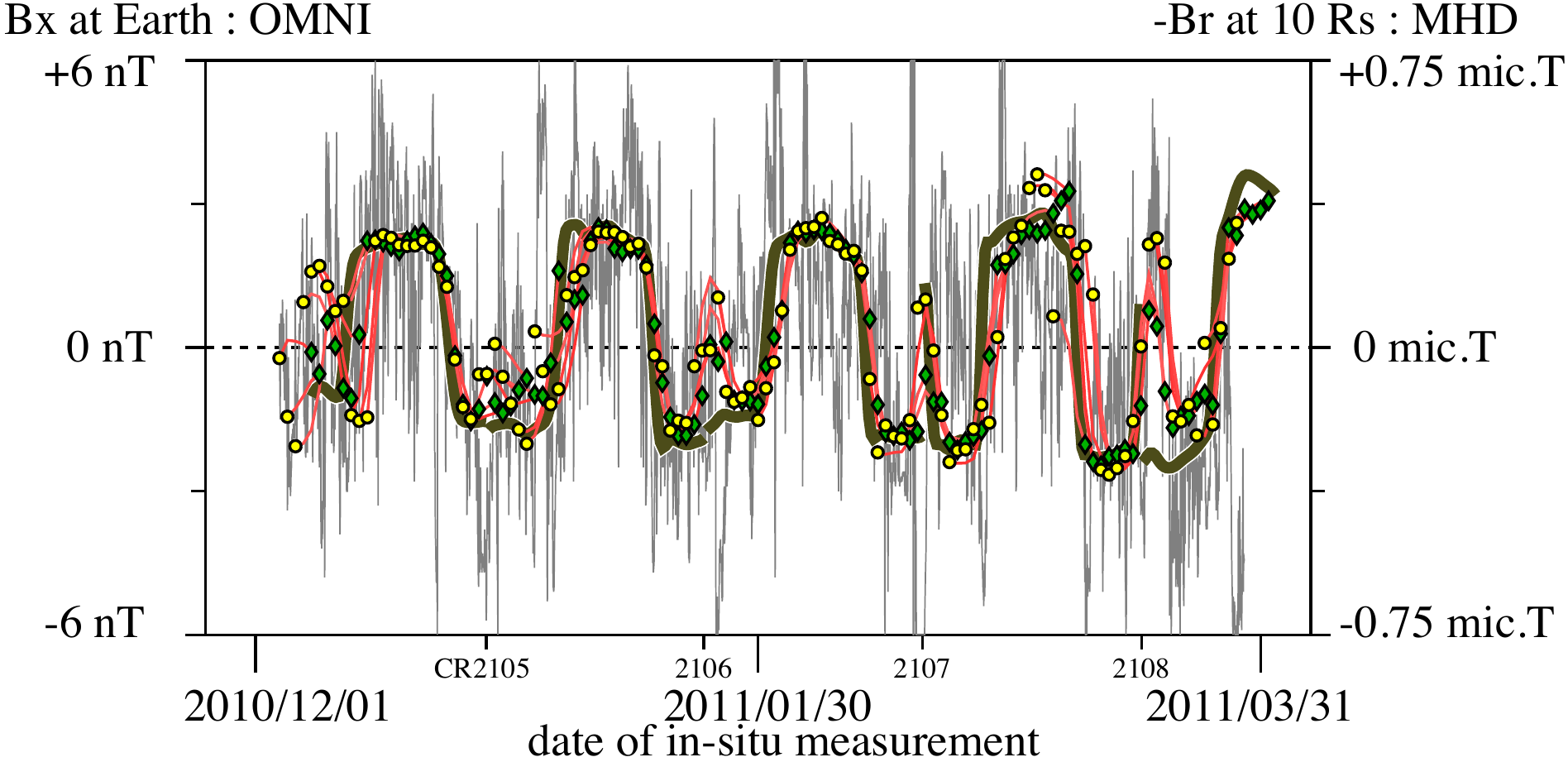}
}
\centerline{
 \includegraphics[width=1.0\textwidth,clip=]{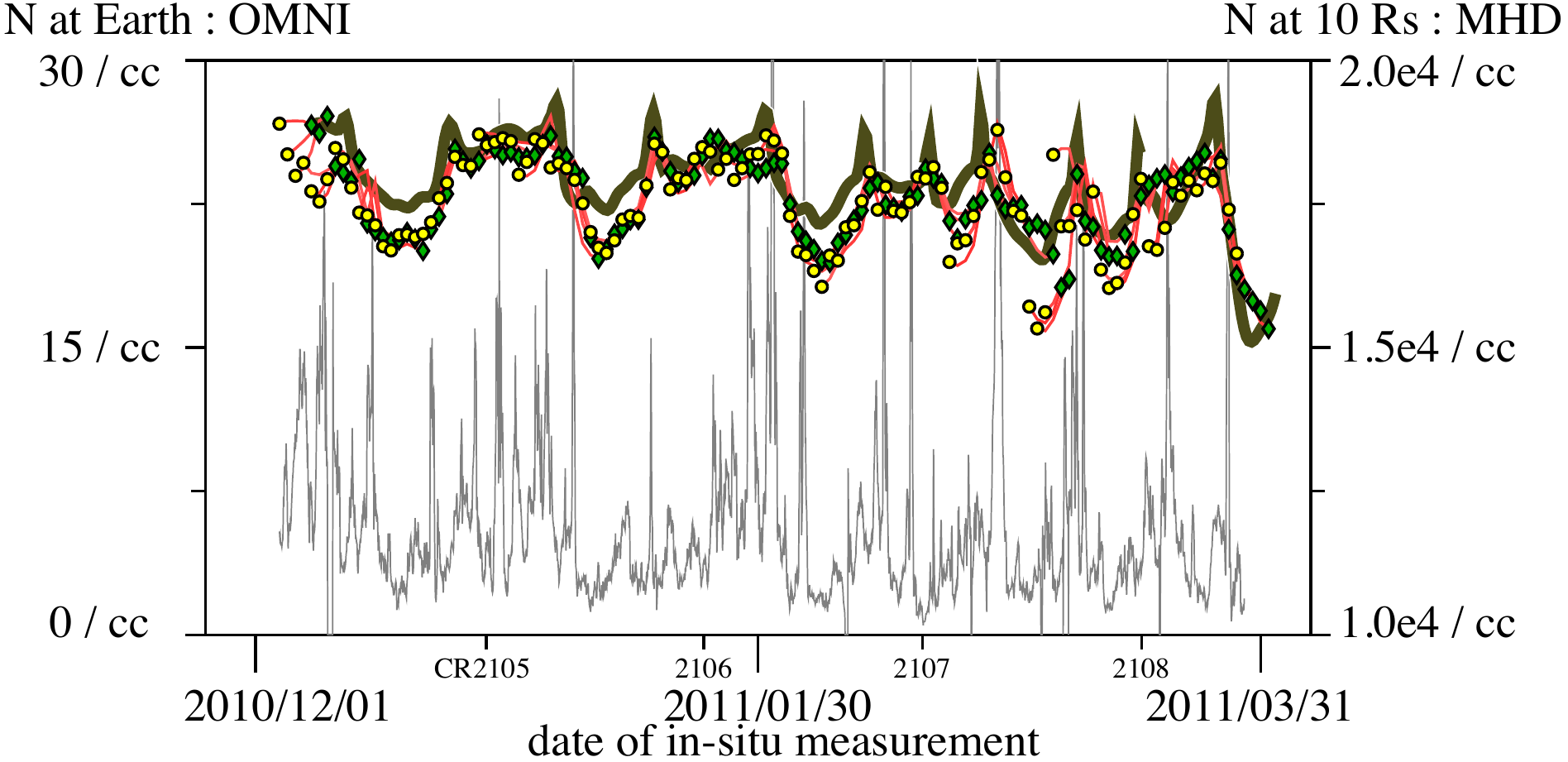}
}
\centerline{
 \includegraphics[width=1.0\textwidth,clip=]{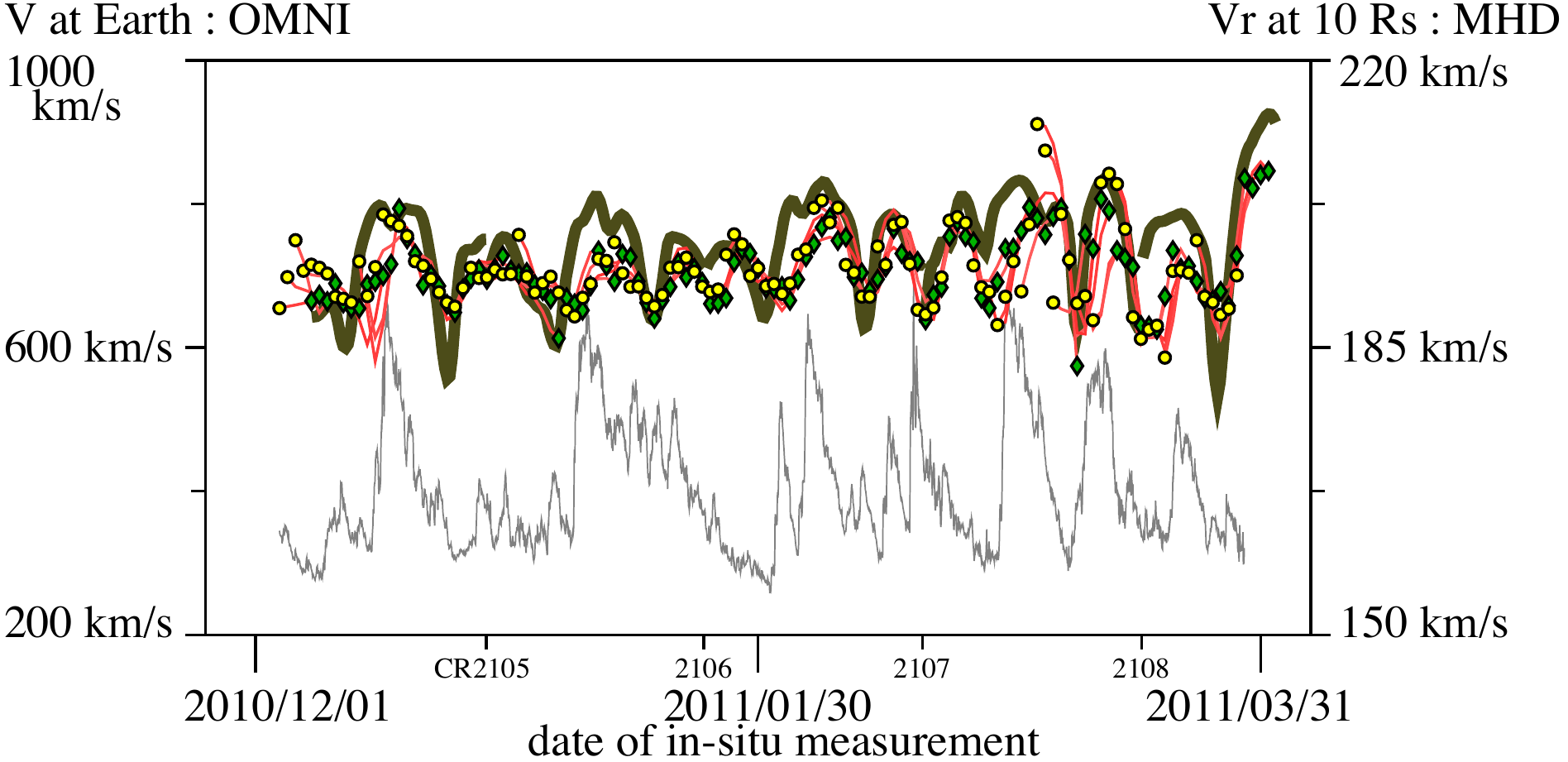}
}
\caption{%
Comparisons of the Sun--Earth component of the interplanetary magnetic field, plasma number density and flow speed, from 1 December 2010 to 31 March 2011.
Gray lines show the hourly averaged \myinsitu data, from OMNIWeb database.
The green-filled diamonds (yellow-filled circles) indicate the values obtained from the daily MHD simulation (B) sampled at the heliographic position of the Earth at noon UT, four days after (the same date as) the corresponding \myinsitu measurement, at 10 ${\rm R}_\mysun$.
The red lines connecting the two circles show the values at the noon of the three days in between in the same daily simulation data.
Dark-brown lines show the simulated values from the medium-resolution simulation (A), prepared in the same manner as the green-filled diamond marks.
Short vertical lines with Carrington rotation numbers are placed at the dates four days earlier than the start date of each rotation period.
}%
\label{fig7}
\end{figure}

Here, we use the simulated $B_r$ at the heliocentric distance of ten ${\rm R}_\mysun$ to compare with the \myinsitu data.
The top plot of Figure \ref{fig7} shows the Sun--Earth component of the interplanetary magnetic field (IMF), that is, $B_x$ from the OMNI database and the corresponding simulation parameter, $[-B_r]$.
In the plot, the hourly-averaged $B_x$ in the OMNI database is drawn with thin gray lines.
The green-filled diamonds and yellow-filled circles show the value of simulated $[-B_r]$.
The green diamond is placed, in the plot, at the position four days after the date for which the daily simulation (B) was computed.
The yellow circle is placed at the date of the \myinsitu measurement and the input magnetic-field map, but the value is sampled at about 53 degrees West of the Sun--Earth direction.
A red curve connects the circle and diamond and shows the values on the three days in between.
A green-filled diamonds and red line show, in this way, a prediction for four days.
Thick dark brown curves show the same quantity as the green-filled diamond derived from the once-per-CR medium-resolution simulations (A).
Overall, good agreement of the simulation data with the \myinsitu near-Earth measurement is obtained.

The once-per-CR simulation (A) is not very good at reproducing the data near 0 or 360 degrees Carrington longitude because of the discontinuity of the input solar photospheric magnetic-field maps.
On the other hand, the daily simulation (B) gives better results around the start or end date of a Carrington rotation period.
We will be able to improve the results of the higher-resolution simulation by conducting it more frequently, for example, twice per a CR rotation using an extra synoptic map from 180 degrees Carrington longitude to 180 degrees of the next rotation period.

The yellow-filled circle points, which represent the now-cast, often trace some peak values of the \myinsitu data better than the green-filled diamond points that are derived from simulation using the magnetic-field data four days earlier.
However, overall, both agree moderately well with the \myinsitu data, and here we refrain from attempting to judge which simulation data value is better than the other.
Instead, we point out here that the solar wind can be often affected by the condition of the photospheric magnetic field at a distant solar surface region, for example about 90 degrees East from its origin.
In the daily simulation (B), the solar-surface region 37 degrees East from the Sun--Earth direction (90 degrees East from the assumed origin) can be included in the daily updated synoptic map made for the day of the \myinsitu measurement but had not been included in the map made four days earlier.
%

In the plot, the scale for the simulated $-B_r$, $\pm 0.75$ $\mu$ (micro) Tesla corresponds to, if it decreases with $1/r^2$, $1.62(=750/21.5^2)$ nT at 1 AU, which is about quarter of the scale used for the \myinsitu data, 6 nT.
The deficiency of the magnetic-field strength derived from the coronal models, so-called missing flux, is an interesting and unsolved topic in field of the solar wind and interplanetary dynamics.
Because there is no consensus about the causes of the mismatch, we do not apply any adjustment, for example multiplying values of $B_r$ by a factor four, to the input magnetic-field map or simulated data.
As a data pipeline product, the MHD simulation data and the magnetic-field maps are published as is.

The middle and bottom panels of Figure \ref{fig7} compare the simulated plasma number density and flow speed at ten solar radii with the OMNI data, respectively.
In these plots, different offsets and scales are applied to enhance small contrasts of these plasma quantities simulated with polytropic assumption.
In the middle plot, many of the maxima and minima of the simulated density coincide with those of the 1-AU measurements.
The correlation coefficients are reasonably good: about 0.35.
Although the contrast of the simulated density is much smaller than those in the \myinsitu measurement data, a key feature of the solar corona, the enhancements of the simulated density at the closed field streamer or pseudo-streamer, can be reasonably reproduced.
In the bottom plot, the flow speed greater than 500 \mykmps\, in the \myinsitu data and the maxima of the simulated plasma flow speed coincide well, although the contrast of the simulated flow speed is smaller.
The simulated higher speed plasma flow originates from the open-field coronal hole.
Therefore, these reasonable agreements in plasma parameters imply that the simulated coronal magnetic field, both open- and closed-field structures, is reasonable overall.

We note that the mass flux of the simulated solar wind tends to be much larger than that measured at Earth.
In the four-month period shown in Figure \ref{fig7}, the mass flux of the simulated solar wind near the solar Equator at ten solar radii is about $3.5\times 10^{6}$ \mykmps \mypcc, and the average of the mass flux measured near the Earth is about $2\times 10^{3}$ \mykmps \mypcc.
Hence the simulated mass flux is about four times as much as in reality (($3.5\times 10^6)/(2\times 10^3)\div (215/10)^2 \approx 3.79$).
The surplus can be reduced by applying the boundary conditions that are capable of limiting the boundary mass flux (\opencite{Hayashi05}).
As a standard data pipeline product, however, the MHD module in the JSOC uses a simple boundary condition without any mass-flux limits on the solar coronal base.

\section{Summary and Discussion}\label{SCTsmmry}
This article described the MHD simulations that are part of the JSOC data pipeline and the specifications of the output data at the JSOC database interface.

Three types of the MHD simulations, labeled (A) to (C) in this paper, are conducted regularly in the HMI data pipeline.
A medium-resolution simulation conducted once per one Carrington rotation period (labeled (A)) and two versions of the daily MHD simulation: One uses the NRT data and the other uses the definitive, calibrated synchronic map data (labeled (B)).
The daily NRT simulations labeled (C) are primarily intended to provide the community with quick-look data to help grasp the current state of the plasma quantities of the solar corona, and the definitive daily simulations provide reliable solutions and supplement the NRT simulations.
For this purpose, we made some compromises in the simulation setting to minimize the computational resources required and accomplish timely productions regularly.

The settings that we chose for making the global magnetic-field map and the MHD simulation of the corona are optimized overall in the context of the HMI data pipeline to provide the simulation data in a timely manner.
The simulation results are compared with the AIA image data and the OMNI near-Earth \myinsitu measurement data, and reasonably good agreements are obtained, meaning the selected parameters are reasonable.

There are three options and functionalities that we did not use in the current version but would include in future.

The first one is to make a polar field correction when constructing the photospheric $B_r$ map.
We have developed a method and tested it with SOHO/MDI data (\opencite{Sun11}) that uses temporal and spatial polynomial interpolation to fill the missing data at the poles.
This polar field correction needs observational data over three years to yield reliable temporal interpolation and one year of data to validate, which was fulfilled by the end of 2014.
The JSOC data pipeline started making synoptic maps and synchronic maps using HMI vector data.
Using the vector-based $B_r$ maps for modeling the solar corona (\eegg \opencite{HayashiJPC13}) is an extra option to improve the input maps.

The second option is in treating the active regions near the longitudinal edges of the whole-surface maps.
In making daily synchronic frames and once-per-CR synoptic maps, it is often the case that only part of the sunspot group is included in the whole-surface map, which can result in substantial imbalance of the magnetic flux and hence displacement of the magnetic neutral lines in the offset map.
Thus, making an improved global map needs the algorithm to recognize magnetically active regions on the solar surface.
It is a reasonable expectation that the HMI Active Region Patches (HARP) module (see \opencite{Bobra14} and \opencite{Turmon15}), already implemented in the HMI vector magnetic data production for automatically identifying magnetically active regions, can improve global map production (\opencite{HayashiSPD12}; \opencite{Hoeksema14}).
These two features, if implemented, will help expand the ways of using whole-surface maps in the models for the global corona such as MHD and PFSS.

The third point is to apply time-dependent boundary conditions to the MHD simulation.
Because the solar corona is a sub-\myAlfvenic system, the state of the solar corona at an instant is influenced by, not only the boundary values at that instant but also those in the past.
Therefore, it is desirable for the MHD model to be capable of handling time-dependent boundary data.
The global variation of the magnetic field, caused by the differential longitudinal flow and the meridional flow, can produce the formation and evolution of large-scale twisted structures that can erupt into interplanetary space (\eegg \opencite{Linker01}; \opencite{Yeates08}; \opencite{Yeates12}; \opencite{YangFeng12}; \opencite{Feng12}; \opencite{Hayashi13}).
A similar strategy can be applied to interplanetary space (\eegg \opencite{Hayashi12}) and the active regions (\eegg \opencite{Cheung12}; \opencite{Jiang13}; \opencite{Inoue13}).
Because implementing the boundary treatment used in these models, at least ours (\opencite{Hayashi12,Hayashi13}), is not computationally expensive, it will not be a hard task to develop a new version of the MHD modules for the HMI data pipeline.
If successful, the new module will be able to provide MHD solutions of the global solar corona that evolves seamlessly in time.

The codes for these three items were already used in other modules in the HMI pipeline or else by the HMI team.
The whole-surface magnetic-field map and the MHD results with the new choices will be published as new series of data at JSOC database, and the current version will continue to be generated.

%
%
\clearpage

\clearpage

\clearpage

\clearpage

\clearpage

\clearpage

\clearpage

%
%
\clearpage
\begin{table}
\caption{Series Names of Synoptic and Synchronic Maps}
\label{tbl1}
\begin{tabular}{ll}
\hline
Series name & Description \\
\hline
\textsf{hmi.Synoptic\_Mr\_720s}* & Radial-component standard synoptic map \\
\textsf{hmi.Synoptic\_Ml\_720s}  &  Line-of-sight component standard synoptic map \\
\textsf{hmi.Mrdailysynframe\_720s}* & Radial-component daily updated map \\
\textsf{hmi.Mldailysynframe\_720s}  & Line-of-sight component daily updated map \\
\textsf{hmi.Mrdailysynframe\_720s\_nrt}* & Radial-component NRT daily updated map \\
\textsf{hmi.Mldailysynframe\_720s\_nrt}  & Line-of-sight component NRT daily updated map\\
\hline
\end{tabular}
\end{table}

\begin{table}
\caption{Parameters in MHD Simulation}
\label{tbl2}
\begin{tabular}{lll}
\hline
Notation & Value &    \\
\hline
$\gamma$& 1.05 & Specific heat ratio \\
$n_\mysubb$ & $2\times 10^8$ count \mypcc & Density at bottom (at $1.01 R_\mysun$)\\
$T_\mysubc$ & 1 M\mykelvin & Temperature at critical point (at $\approx 5.5 R_\mysun$) \\
$\Omega$ & 360 degree $/$ 25.3 days & Sidereal solar rotation rate \\
\hline
\end{tabular}
\end{table}

\begin{table}
\caption{Summary of the MHD runs and the input HMI magnetogram map data}
\label{tbl3}
\begin{tabular}{rlll}
\hline
Label & Output series name & input series name & cadence \\
\hline
a&\textsf{hmi.MHDcorona}            &\textsf{hmi.Synoptic\_Mr\_720s}        & 27 or 28 days \\
 &                                  &                                       & (1 Carrington Rot.) \\
b&\textsf{hmi.MHDcorona\_daily}     &\textsf{hmi.Mrdailysynframe\_720s}     & 1 day \\
c&\textsf{hmi.MHDcorona\_daily\_nrt}&\textsf{hmi.Mrdailysynframe\_720s\_nrt}& 1 day \\
\hline
\end{tabular}
\end{table}

\begin{table}
\caption{Segments of coronal MHD data cube}
\label{tbl4}
\begin{tabular}{llcl}
\hline
Segment name & Description & Notation & Unit \\
\hline
\textsf{N} & Number density & $N$ & $10^8$\mypcc\\
\textsf{T} & Temperature    & $T$ & $10^6$\mykelvin \\
\textsf{Vr}& Radial component of plasma flow speed      &$V_r$     &\mykmps\\
\textsf{Vt}& Latitudinal component of plasma flow speed &$V_\theta$&\mykmps\\
\textsf{Vp}& Longitudinal component of plasma flow speed&$V_\phi$  &\mykmps\\
\textsf{Br}& Radial component of magnetic field      &$B_r$     &Gauss\\
\textsf{Bt}& Latitudinal component of magnetic field &$B_\theta$&Gauss\\
\textsf{Bp}& Longitudinal component of magnetic field&$B_\phi$  &Gauss\\
\hline
\end{tabular}
\end{table}

\clearpage

\begin{landscape}

\begin{table}
\caption{%
Selected keywords with the same value as those of the input data.
In the rightmost three columns, marks `x' indicates that the output MHD data by (A) moderate-resolution once-per-CR simulation using definitive synoptic map, (B) low-resolution simulation using daily definitive synchronic map, or (C) low-resolution simulation using daily NRT synchronic map has the keyword.
The capital `X' indicates that the keyword is the prime keyword used to identify the data record: The MHD data series and the magnetogram map series share the same prime keyword.
}%
\label{tbl5}
\begin{tabular}{lllccc}
\hline
name   & type,value,[unit] & description & (A) & (B) & (C) \\
\hline
\textsf{T\_REC}    & TAI time format & Time of Record          &   & X & X \\
\textsf{CAR\_ROT}  & 4-digit integer & Carr. Rot. Num. of obs. & X & x & x \\
\hline
\textsf{CARRTIME} & [degree] & Carrington Time in float        & x & x & x \\
\textsf{MAP\_DATE} & ISO 8601 format UTC & date of input map creation & x & x & x \\
\textsf{MAP\_CVER} & string & Code version info. of synoptic/synchronic map module & x & x & x \\
\textsf{MAP\_BLDV} & string & Build number of synoptic map module   & x & x & x \\
\textsf{CADENCE} &  86 400 & Cadence of input data [seconds]& x &   &   \\
                 &     360 &                       [degrees]&   & x & x \\
\textsf{T\_OBS}        & TAI time format & nominal time of observation &  & x & x \\
\textsf{T\_REC\_epoch} & TAI time format & Time of origin & & x & x \\
\textsf{T\_REC\_step} & 720    & \textsf{ts\_eq} step & & x & x \\
\textsf{T\_REC\_unit} & `secs' & \textsf{ts\_eq} unit & & x & x \\
\textsf{T\_START}  & TAI time format & Carrington Rotation Start Time of input map & x & & \\
\textsf{T\_STOP}   & TAI time format & Carrington Rotation Stop Time of input map  & x & & \\
\textsf{T\_ROT}    & TAI time format & Carrington Rotation Middle Time of input map& x & & \\
\textsf{B0\_ROT}   & [degree] & B-zero angle ($B_\mysun$) of map center& x & & \\
\textsf{B0\_FRST}  & [degree] & B-zero angle ($B_\mysun$) at \textsf{T\_START} & x & & \\
\textsf{B0\_LAST}  & [degree] & B-zero angle ($B_\mysun$) at \textsf{T\_LAST}  & x & & \\
\textsf{LON\_FRST} & [degree] & First Carrington Time of input global map & x & x & x \\
\textsf{LON\_LAST} & [degree] & Last Carrington Time of input global map  & x & x & x \\
\textsf{ORIGIN}   & `SDO/JSOC-SDP' & ORIGIN: location where file was made & x & x & x \\
\textsf{TELESCOP} & `SDO/HMI'      & For HMI: \textsf{SDO/HMI} & x & x & x \\
\textsf{INSTRUME} & `HMI\_SIDE1'   & For HMI: \textsf{HMI\_}\{\textsf{SIDE1}, \textsf{FRONT2}, \textsf{COMBINED}\} & x & x & x \\
\textsf{CALVER64} & hex & Calibration Version & x & x & x \\
\hline
\textsf{COMMENT}  & & Supplemental Comments & x & x & x \\
\hline
\end{tabular}
\end{table}

\end{landscape}
\clearpage
\begin{landscape}
  
\begin{table}
\caption{%
JSOC keywords of the MHD data: new or updated by the MHD modules.
Asterisks (*) are placed next to the keywords newly added by the MHD modules.
Keywords without the asterisks in this table are of the same name as those of the input magnetic-field maps but given different values.
}%
\label{tbl6}
\begin{tabular}{lllccc}
\hline
name   & type,value,[unit] & description & (A) & (B) & (C) \\
\hline
\textsf{MHD\_VER1} *& string & version of JSOC interface code & x & x & x \\
\textsf{MHD\_VER2} *& string & version of MHD code & x & x & x \\
\textsf{MHD\_SET1} *& string & setting in coronal MHD model & x & x & x \\
\textsf{MHD\_SET2} *& string & setting in coronal MHD model & x & x & x \\
\textsf{MHD\_SET3} *& string & setting in coronal MHD model (for future use)& x & x & x \\
\textsf{MHDIBMAG}  *& string & initial and boundary mag. setting & x & x & x \\
\textsf{MHDMGIDX}  *& integer & index of initial mag. setting & x & x & x \\
\textsf{MHDMODEL}  *& string & supplemental comment about model & x & x & x \\
\textsf{INPUTMAP}  *& string & input magnetic global map identifier& x & x & x \\
\hline
\textsf{DATE}    & ISO 8601 format UTC & time of processing  & x & x & x \\
\textsf{CRPIX1}  & 72.5 & location of the image center & x &   &   \\
                 & 36.5 &                              &   & x & x \\
\textsf{CRPIX2}  & 36.5 & location of the image center & x &   &   \\
                 & 18.5 &                              &   & x & x \\
\textsf{CRPIX3} *& 40.5 & location of the image center & x & x & x \\
\textsf{CTYPE1}  & `CARL-CAR' & Carrington Time                 & x & x & x \\
\textsf{CTYPE2}  & `CRLT-CAR' & Heliographic latitude           & x & x & x \\
\textsf{CTYPE3} *& `HECR'     & Radial distance from sun center & x & x & x \\
\textsf{CRVAL1}  & [degree]   & Carrington time at center of the data       & x & x & x \\
\textsf{CRVAL2}  &  0.0 & Latitude at center of the data              & x & x & x \\
\textsf{CRVAL3} *&  3.0 & Heliocentric distance at center of the data & x & x & x \\
\textsf{CDELT1}  & -2.5 & image scale in the 1st direction (Carrington time) & x &   &   \\
                 & -5.0 &                                                    &   & x & x \\
\textsf{CDELT2}  &  2.5 & image scale in the 2nd direction (latitude) & x &   &   \\
                 &  2.5 &                                             &   & x & x \\
\textsf{CDELT3} *& 0.05 & image scale in the 3rd direction (radius)   & x & x & x \\
\textsf{CUNIT1}  & `degree' & UNIT of scale in the 1st direction & x & x & x \\
\textsf{CUNIT2}  & `degree' & UNIT of scale in the 2nd direction & x & x & x \\
\textsf{CUNIT3} *& `solRad' & UNIT of scale in the 3rd direction & x & x & x \\
\textsf{WCSNAME} & `3D-SPHERICAL' & WCS system name & x & x & x \\
\hline
\end{tabular}
\end{table}

\end{landscape}

%
%
%
%
\begin{acks}%
We thank the many team members who have contributed to the success of the SDO
mission and particularly to the HMI instrument.
This work was supported by NASA Contract NAS5-02139 (HMI) to Stanford University.
We use the OMNIweb database at GSFC of NASA and the SDO/AIA image data at LMSAL.
\end{acks}
\section*{Disclosure of Potential Conflicts of Interest}
The authors declare that they have no conflicts of interest.
\end{article} 

\begin{thebibliography}{}
\bibitem[\protect\citeauthoryear{{Altschuler and Newkirk}}{1969}]{Altschuler69} Altschuler,~M.D., Newkirk,~Jr.G.: 1969, \solphys{} \textbf{9}, 131.
\bibitem[\protect\citeauthoryear{{Bobra \etal}}{2014}]{Bobra14} Bobra,~M., Sun,~X., Hoeksema,~J.T., Turmon,~M., Liu,~Y., Hayashi,~K., et al.: 2014, \solphys{} \textbf{289}, 3549.
\bibitem[\protect\citeauthoryear{{Brackbill and Barnes}}{1980}]{Brackbill80} Brackbill,~J.U., Barnes,~D.C.: 1980, \jcp{} \textbf{35}, 426.
\bibitem[\protect\citeauthoryear{{Brio and Wu}}{1988}]{Brio88} Brio,~M., Wu,~C.C.: 1988, \jcp{} \textbf{75}, 400.
\bibitem[\protect\citeauthoryear{{Centeno \etal}}{2014}]{Centeno14} Centeno,~R., Schou,~J., Hayashi,~K., Norton,~A., Hoeksema,~J.T., Liu,~Y., et al.: 2014, \solphys{} \textbf{289}, 3531.
\bibitem[\protect\citeauthoryear{{Cheung and DeRosa}}{2012}]{Cheung12} Cheung,~M.C.M., DeRosa,~M.L.: 2012, \apj{} \textbf{757}, 147.
\bibitem[\protect\citeauthoryear{{Dryer \etal}}{1991}]{Dryer91} Dryer,~M., Smith,~Z.K., Coates,~A.J., Johnstone,~A.D.: 1991, \solphys{} \textbf{132}, 353.
\bibitem[\protect\citeauthoryear{{Feng \etal}}{2010}]{Feng10} Feng,~X.S., Yang,~L.P., Xiang,~C.Q., Wu,~S.T., Zhou,~Y.F., Zhong,~D.K.: 2010, \apj{} \textbf{723}, 300.
\bibitem[\protect\citeauthoryear{{Feng \etal}}{2012}]{Feng12} Feng,~X.S., Jiang,~C., Changqing,~X., Zhao,~X.P., Wu,~S.T.: 2012, \apj{} \textbf{758}, 62.
\bibitem[\protect\citeauthoryear{{Han, Wu, and Dryer}}{1988}]{Han88} Han,~S.M., Wu,~S.T., Dryer,~M.: 1988, \textit{Comp. Fluids}{} \textbf{16}, 81.
\bibitem[\protect\citeauthoryear{{Hayashi}}{2005}]{Hayashi05} Hayashi,~K.: 2005, \apjss{} \textbf{161}, 480.
\bibitem[\protect\citeauthoryear{{Hayashi}}{2012}]{Hayashi12} Hayashi,~K.: 2012, \jgr{} \textbf{117}, A08105.
\bibitem[\protect\citeauthoryear{{Hayashi}}{2013}]{Hayashi13} Hayashi,~K.: 2013, \jgr{} \textbf{118}, 6889.
\bibitem[\protect\citeauthoryear{{Hayashi, Zhao, and Liu}}{2008}]{Hayashi08} Hayashi,~K., Zhao,~X.P., Liu,~Y.: 2008, \jgr{} \textbf{113}, A07104.
\bibitem[\protect\citeauthoryear{{Hayashi \etal}}{2012}]{HayashiSPD12} Hayashi,~K., Liu,~Y., Sun,~X., Turmon,~M.J.: 2012, SPD 43rd / AAS 220th meeting 2012 (\#207.16); BAAS vol.44.
\bibitem[\protect\citeauthoryear{{Hayashi \etal}}{2013}]{HayashiJPC13} Hayashi,~K., Hoeksema,~J.T., Liu,~Y., Norton,~A., Sun,~X., Centeno,~R., et al.: 2013, Proceedings of GONG2012/LWS/SDO5/SOHO27, Eclipse on the Coral Sea, Cycle 24 Ascending, \textit{J. Phys. Conf. Ser.}{} \textbf{440}, 012036.
\bibitem[\protect\citeauthoryear{{Hoeksema \etal}}{2014}]{Hoeksema14} Hoeksema,~J.T., Liu,~Y., Hayashi,~K., Sun,~X., Schou,~J., Couvidat,~S., et al.: 2014, \solphys{} \textbf{289}, 3483.
\bibitem[\protect\citeauthoryear{{Inoue \etal}}{2013}]{Inoue13} Inoue,~S., Hayashi,~K., Magara,~T., Choe,~G.S., Park,~Y.D.: 2013, \apj{} \textbf{788}, 182.
\bibitem[\protect\citeauthoryear{{Jiang \etal}}{2013}]{Jiang13} Jiang,~C., Feng,~X.S., Wu,~S.T., Hu,~Q.: 2013, \apjl{} \textbf{771}, 30.
\bibitem[\protect\citeauthoryear{{Lemen \etal}}{2012}]{Lemen12} Lemen,~J.R., Title,~A.M, Akin,~D.J., Boerner,~P.F., Chou,~C., Drake,~J.F., et al.: 2012, \solphys{} \textbf{275}, 17.
\bibitem[\protect\citeauthoryear{{Linker \etal}}{1990}]{Linker90} Linker,~J.A., van Hoven,~G., Schnack,~D.D.: 1990, \grl{} \textbf{17}, 2281.
\bibitem[\protect\citeauthoryear{{Linker \etal}}{2001}]{Linker01} Linker,~J.A., Lionello,~R., Mikic,~Z., Amari~T.: 2001, \jgr{} \textbf{106}, 25165.
\bibitem[\protect\citeauthoryear{{Liu \etal}}{2012}]{Liu12} Liu,~Y., Hoeksema,~J.T., Scherrer,~P.H., Schou,~J., Couvidat,~S., Bush,~R.I., et al.: 2012, \solphys{} \textbf{279}, 295.
\bibitem[\protect\citeauthoryear{{Mikic \etal}}{1999}]{Mikic99} Mikic,~Z., Linker,~J.A., Schnack,~D.D., Lionello,~R., Tarditi,~A.: 1999, \textit{Phys. Plasma}{} \textbf{6}, 2217.
\bibitem[\protect\citeauthoryear{{Nakagawa \etal}}{1987}]{Nakagawa87} Nakagawa,~Y., Hu,~Y.Q., Wu,~S.T.: 1987, \aap{} \textbf{197}, 354.
\bibitem[\protect\citeauthoryear{{Nakamizo \etal}}{2009}]{Nakamizo09} Nakamizo,~A., Tanaka,~T., Kubo,~Y., Kamae,~S., Shimizu,~H.: 2009, \jgr{} \textbf{114}, A07109.
\bibitem[\protect\citeauthoryear{{Parker}}{1958}]{Parker58} Parker,~E.N.: 1958, \apj{} \textbf{128}, 664.
\bibitem[\protect\citeauthoryear{{Powell}}{1994}]{Powell94} Powell,~K.G.: 1994, {\it ICASE Report}, 94-24, NASA Langley Research Center: Hampton, Virginia
\bibitem[\protect\citeauthoryear{{Powell et al.}}{1999}]{Powell99} Powell,~K.G., Roea~,P.L., Linde,~T., Gombosi,~T.I., de Zeeuw,~D.L.: 1999, \jcp{} \textbf{154}, 284.
\bibitem[\protect\citeauthoryear{{Roe and Balsara}}{1996}]{Roe96} Roe,~P.L., Balsara,~D.S.: 1996, SIAM, J. Appl. Math.{} \textbf{56}, 57.
\bibitem[\protect\citeauthoryear{{Schatten, Wilcox, and Ness}}{1969}]{Schatten69} Schatten,~K.H., Wilcox,~J.M., Ness,~N.F.: 1969, \solphys{} \textbf{6}, 442.
\bibitem[\protect\citeauthoryear{{Scherrer \etal}}{2012}]{Scherrer12} Scherrer,~P., Schou,~J., Bush,~R.I., Kosovichev,~A.G., Bogart,~R.S., Hoeksema,~J.T., et al.: 2012, \solphys{} \textbf{275}, 207.
\bibitem[\protect\citeauthoryear{{Schou \etal}}{2012}]{Schou12} Schou,~J., Scherrer,~P.H., Bush,~R.I., Wachter,~R., Couvidat,~S., Rabello-Soares,~M.C., et al.: 2012, \solphys{} \textbf{275}, 229.
\bibitem[\protect\citeauthoryear{{Suess \etal}}{1995}]{Suess95} Suess,~S.T., Wang,~A.-H., Wu,~S.T.: 1995, \jgr{} \textbf{101}, 19957.
\bibitem[\protect\citeauthoryear{{Sun \etal}}{2011}]{Sun11} Sun,~X., Liu,~Y., Hoeksema,~J.T, Hayashi,~K., Zhao,~X.P.: 2011, \solphys{} \textbf{270}, 9.
\bibitem[\protect\citeauthoryear{{Tanaka}}{1995}]{Tanaka95} Tanaka,~T.: 1995, \jgr{} \textbf{100}, 12057.
\bibitem[\protect\citeauthoryear{{Toth}}{2000}]{Toth00} Toth,~G.: 2000, \jcp{} \textbf{161}, 605.
\bibitem[\protect\citeauthoryear{{Turmon \etal}}{2015}]{Turmon15} Turmon,~M., Hoeksema,~J.T., Sun,~X., Bobra,~M., Sommers,~J.: 2015, {\it The Helioseismic and Magnetic Imager (HMI) magnetic field pipeline: HMI active region patches}, \solphys, in preparation
\bibitem[\protect\citeauthoryear{{Ulrich \etal}}{2002}]{Ulrich02} Ulrich,~R.K., Schott,~E., Boyden,~J.E., Webster,~L.: 2002, \apjss{} \textbf{139}, 259.
\bibitem[\protect\citeauthoryear{{Usmanov}}{1993}]{Usmanov93} Usmanov,~A.V.: 1993, \solphys{} \textbf{146}, 377.
\bibitem[\protect\citeauthoryear{{Usmanov and Dryer}}{1995}]{Usmanov95} Usmanov,~A.V., Dryer,~M.: 1995, \solphys{} \textbf{159}, 347.
\bibitem[\protect\citeauthoryear{{Usmanov \etal}}{2000}]{Usmanov00} Usmanov,~A.V., Goldstein,~M.L., Besser,~B.P., Fritzer,~J.M.: 2000, \jgr{} \textbf{105}, 12675.
\bibitem[\protect\citeauthoryear{{Usmanov \etal}}{2011}]{Usmanov11} Usmanov,~A.V., Matthaeus,~W.H., Breech,~B.A., Goldstein,~M.L.: 2011, \apj{} \textbf{727}, 84.
\bibitem[\protect\citeauthoryear{{van Leer}}{1979}]{vanLeer79} van Leer,~B.: 1979, \jcp{} \textbf{32}, 101.
\bibitem[\protect\citeauthoryear{{Wang and Sheeley}}{1992}]{Wang92} Wang,~Y.-M., Sheeley,~N.R.~Jr.: 1992, \apj{} \textbf{392}, 310.
\bibitem[\protect\citeauthoryear{{Washimi and Sakurai}}{1993}]{Washimi93} Washimi,~H., Sakurai,~T.: 1993, \solphys{} \textbf{143}, 173.
\bibitem[\protect\citeauthoryear{{Wu and Wang}}{1987}]{Wu87} Wu,~S.T., Wang,~J.F.: 1987, \textit{Comp. Meth. Appl. Mech. Eng.}{} \textbf{64}, 267.
\bibitem[\protect\citeauthoryear{{Wu, Dryer, and Han}}{1983}]{Wu83} Wu,~S.T., Dryer,~M., Han,~S.M.: 1983, \solphys{} \textbf{84}, 395.
\bibitem[\protect\citeauthoryear{{Yang \etal}}{2012}]{YangFeng12} Yang,~L.P., Feng,~X.S., Xiang,~C.Q, Liu,~Y., Zhao,~X.P., Wu,~S.T.: 2012, \jgr{} \textbf{117}, A08110.
\bibitem[\protect\citeauthoryear{{Yeates and Mackay}}{2012}]{Yeates12} Yeates,~A.R., Mackay,~D.H.: 2012, \apjl{} \textbf{753}, 34.
\bibitem[\protect\citeauthoryear{{Yeates, Mackay, and van Ballegooigen}}{2008}]{Yeates08} Yeates,~A.R., Mackay,~D.H., van Ballegooijen,~A.: 2008, \apj{} \textbf{680}, 165.
\bibitem[\protect\citeauthoryear{{Yeh and Dryer}}{1985}]{Yeh85} Yeh,~T., Dryer,~M.: 1985, \apss{} \textbf{117}, 165.
\bibitem[\protect\citeauthoryear{{Zhao, Hoeksema, Scherrer}}{1999}]{Zhao99} Zhao,~X.P., Hoeksema,~J.T., Scherrer,~P.H.: 1999, \jgr{} \textbf{104}, 9735.
\bibitem[\protect\citeauthoryear{{Zhao and Webb}}{2003}]{Zhao03} Zhao,~X.P., Webb,~D.F.: 2003, \jgr{} \textbf{108}, 1234.
\bibitem[\protect\citeauthoryear{{Zhao \etal}}{2012}]{Zhao12} Zhao,~J., Couvidat,~S., Bogart,~R.S., Parchevsky,~K.V., Birch,~A.C., Duvall,~T.L., et al.: 2012, \solphys{} \textbf{275}, 375.
\end{thebibliography}
\end{document}